\documentclass[aps,prd,twocolumn,superscriptaddress,floatfix,showpacs,showkeys]{revtex4-1}

\usepackage[utf8]{inputenc}
\usepackage[T1]{fontenc}
\usepackage{slashed}
\usepackage{times}
\usepackage{mathrsfs}
\pdfoutput=1
\usepackage{amssymb,amsfonts,amsmath,amsthm,wasysym}
\usepackage{dsfont,bbm}
\usepackage{dcolumn,textcomp}
\usepackage{placeins}

\usepackage{lmodern}
\usepackage[british]{babel}
\usepackage{amsmath,mleftright}
\usepackage{comment}
\usepackage{xparse}
\usepackage{braket}
\usepackage{amssymb}
\usepackage{amsfonts}
\usepackage{color}
\usepackage{float}
\usepackage{stmaryrd}
\usepackage{url}
\usepackage{hyperref}
\usepackage{subfigure}
\usepackage{placeins,bbm}
\usepackage{subfigure}

\usepackage{booktabs,adjustbox} % required by to_latex of pandas
\DeclareOldFontCommand{\bf}{\normalfont\bfseries}{\mathbf}
\usepackage{caption}
\usepackage{slashed}
\usepackage{bbold}

\def\MSbar{\overline{\mathrm{ MS}}}

\NewDocumentCommand{\evalat}{sO{\big}mm}{%
  \IfBooleanTF{#1}
   {\mleft. #3 \mright|_{#4}}
   {#3#2|_{#4}}%
}

\newcommand{\AdjQCD}{AdjQCD}

\begin{document}
\title{Lattice simulations of adjoint QCD with one Dirac overlap fermion}

\author{Georg Bergner\thanks{georg.bergner@uni-jena.de}}
\affiliation{University of Jena, Institute for Theoretical Physics,\\ 
	Max-Wien-Platz 1, D-07743 Jena, Germany}
\affiliation{University of M\"unster, Institute for Theoretical Physics, 
	Wilhelm-Klemm-Str.~9, D-48149 M\"unster, Germany}
\author{Juan Camilo Lopez}
\affiliation{University of Jena, Institute for Theoretical Physics,\\ 
	Max-Wien-Platz 1, D-07743 Jena, Germany}
\author{Stefano Piemonte\thanks{stefano.piemonte@ur.de}}
\affiliation{University of Regensburg, Institute for Theoretical Physics, 
	Universit\"atsstr.~31, D-93040 Regensburg, Germany}
\author{Ivan Soler Calero}
\affiliation{University of Jena, Institute for Theoretical Physics,\\ 
	Max-Wien-Platz 1, D-07743 Jena, Germany}

\date{\today}

\begin{abstract}
In this work we investigate the infrared behaviour of a Yang-Mills theory coupled to a massless fermion in the adjoint representation of the gauge group SU(2). This model has many interesting properties, corresponding to the $\mathcal{N}=2$ Super-Yang-Mills theory without scalars and in the recent years there has been an increasing interest toward understanding whether confinement and fermion condensation occur at low energy. We simulate the theory on the lattice close to the massless limit using the overlap discretization of the fermion action, allowing a precise and clean study of the chiral symmetry breaking pattern and of the fermion condensate. We present results for the scale setting, the condensate and the running of the coupling constant through the gradient flow -- all of them pointing to a theory without an infrared fixed point and remaining confined deep in the infrared regime.
\end{abstract}

\maketitle

\section{Introduction}

Non-Abelian Yang-Mills theories coupled to fermions in the adjoint representation of the gauge group (AdjQCD) have properties similar to ordinary QCD, while featuring many additional symmetries which are absent or broken in gauge theories interacting with fundamental fermions. The most notable examples are supersymmetry, center symmetry and discrete axial symmetry. For instance, the model of strong interactions between a gauge field and a massless adjoint Majorana fermion corresponds to $\mathcal{N}=1$ Super Yang-Mills theory (SYM). The theory with two conserved supercharges, $\mathcal{N}=2$ SYM, can be broken down to a simple gauge theory coupled to an adjoint Dirac fermion, namely $N_f=1$ AdjQCD, if the mass of the scalars is sent to infinity. When also one of the two Majorana components of the Dirac fermion becomes infinitely heavy, $N_f=1$ AdjQCD is further reduced to $\mathcal{N}=1$ SYM and supersymmetry is restored. From this perspective, among all possible models, $N_f=1$ AdjQCD is a simple but yet very interesting model for testing the validity of the Seiberg-Witten electromagnetic duality \cite{SEIBERG199419,Seiberg:1994pq}, being a connecting bridge between two pure supersymmetric theories. We are interested in particular in understanding whether $N_f=1$ AdjQCD has an infrared (IR) fixed point and lies inside the conformal window or it is slightly below the conformal boundary and it presents a dynamically generated scale like a chiral condensate.

If the theory lies below the conformal window, it is also interesting to ask whether confinement and chiral symmetry breaking are both simultaneously present. The interplay between these two non-perturbative phenomena is already alone a good motivation to study $N_f=1$ AdjQCD on the lattice. While the transition from the hadronic matter to the quark-gluon plasma at high temperature is only a smooth cross-over in QCD, if quarks are replaced by adjoint fermions, center symmetry is preserved and a deconfinement phase transition occurs at some critical temperature $T_c$. The anomalous breaking of $U_A(1)$ axial symmetry for adjoint fermions leaves the partition function still invariant under a discrete subgroup of axial rotations, opening therefore the possibility for spontaneous symmetry breaking even with just a single Majorana fermion. \AdjQCD~theories are therefore key models toward the study of how confinement and chiral symmetry breaking are related one another.

If the gauge group is SU($N_c$), adjoint fermions have the property of contributing equally with gluons to the large $N_c$ limit \cite{Azeyanagi:2010ne, Unsal:2010qh}, contrary to fundamental quarks that are going to decouple at leading order when $N_c \rightarrow \infty$. When the number of fermion $N_f$ is too large however, the $\beta$-function can quickly develop an infrared fixed point, where the running of the coupling freezes. In this case the theory is infrared conformal, meaning for instance that the critical deconfinement temperature $T_c$ scales to zero in the fermion massless limit. While $\mathcal{N}=1$ SYM has been proven to be confined and chirally broken \cite{Bergner:2015adz,Ali:2018dnd,Bergner:2015adz}, $N_f=2$ lies already within the conformal window, as probably $N_f=3/2$  \cite{bergner_spectrum_2017,bergner_low_2018,DelDebbio:2009fd,Hietanen:2009az}. Therefore, there is the possibility that $N_f=1$ AdjQCD lies within the lower edge of the conformal window, given the large number of degrees of freedom of adjoint fermions. Indeed, previous lattice investigations are pointing in this direction \cite{Georg:2015,Athenodorou:2021wom,Bi:2019gle}. An infrared fixed point in the running of the coupling would be a striking and surprising discovery, given that $\mathcal{N}=2$ SYM, including even more matter content, is well known to be asymptotically free \cite{SEIBERG199419,Seiberg:1994pq}. This result would imply a very rich phase space connecting $\mathcal{N}=1$ to $\mathcal{N}=2$ SYM when the masses of the scalars and of one of the Majorana fermions are sent to infinity.

't Hooft anomaly matching arguments suggest also a third alternative scenario for the low energy effective theory of $N_f=1$ AdjQCD, where a dynamical scale generation would be provided by a four-fermion condensate in place of a vanishing vacuum expectation value of the standard chiral condensate \cite{Anber:2018iof,Poppitz:2021cxe}. In this case, massless baryons would be required to correctly saturate all anomalies. 

Distinguishing a genuine conformal theory from a confining theory near the lower edge of the conformal window is a challenging task. Non-perturbative lattice simulations can explore a regime where supersymmetry is broken, and in general strong interactions outside the perturbative regime. They are, however, limited to a certain range of scales. In this contribution we will provide strong numerical evidence that the theory has a scale provided by the breaking of chiral symmetry in the range of considered parameters. In section \ref{sec:chiral} we will show the presence of a non-vanishing vacuum expectation value of the chiral condensate. Fermion condensation is already a strong signal for the theory not being infrared conformal, as the chiral condensate provides a natural low-energy scale to the theory. Moreover, in section \ref{sec:flow} we will also study the behaviour of the renormalized gauge coupling thanks to the Wilson flow, providing more evidence of the non-conformality. First, we will show how a non-vanishing scale can be defined through the Wilson flow even when the chiral limit is taken. Finally the running of the strong coupling itself will show no evidence of a fixed point even for energy regions already deep in the infrared regime.

\section{$N_f=1$ adjoint QCD}

\subsection{Continuum action}
In this section we begin by recalling the most important symmetries of the SU(2) gauge theory coupled to one massless fermion in the adjoint representation. The Lagrangian in the continuum reads
\begin{eqnarray}
    \mathcal{L} & = &-\frac{1}{2}\textrm{Tr}(G_{\mu\nu}(x)G^{\mu\nu}(x)) + \nonumber \\ 
    & & \overline{\psi}(x)\gamma^\mu(\partial_\mu+i g A^{a}_\mu(x) T_a^A) \psi(x)\,,
    \label{adjoint_Lagrangian}
\end{eqnarray}
where the field strength tensor $G_{\mu\nu}$ is $\partial_\mu A_\nu-\partial_\nu A_\mu + ig[A_\mu, A_\nu ]$. The generators $T_a^A$ of the gauge group SU(2) act on the Dirac fermion field $\psi$ in the adjoint representation.

As in ordinary QCD, AdjQCD has a conserved $U_V(1)$ vector symmetry and a $U_A(1)$ axial symmetry broken by anomaly. The first difference with respect to QCD is that axial anomaly leaves the partition function invariant under a discrete $Z_{2N_c}$ subgroup. A second difference is a peculiar flavor symmetry appearing already with a single adjoint Dirac fermion. In fact, as a gauge group element in the adjoint representation is real, the real and imaginary parts of the Dirac spinor do not mix, which means it decouples into two Majorana spinors in Minkowski space
\cite{Georg:2015}
\begin{equation}
    \psi = \frac{1}{\sqrt{2}}(\lambda_+ + i\lambda_-)
\end{equation}
where
\begin{equation}
    \lambda_+=\frac{\psi + C\bar{\psi}^T}{\sqrt{2}}, \quad  \lambda_-=\frac{\psi - C\bar{\psi}^T}{\sqrt{2}i}\,.
\end{equation}
The two components fulfill the Majorana condition by construction and they can be combined into a Dirac fermion field $\lambda\equiv(\lambda_+,\lambda_-)$. After this decomposition, the Lagrangian of Eq.~\ref{adjoint_Lagrangian} can be rewritten as
\begin{equation}
    \mathcal{L}=\frac{1}{2}\sum_k\lambda_k(x)(i\slashed{D})\lambda_k(x) -\frac{1}{2}\text{Tr}((G_{\mu\nu}(x)G^{\mu\nu}(x))
    \label{eq: Majorana_Lagrangian}
\end{equation}
where $k=+,-$. A ``two-flavour'' $SU(2)$ symmetry of the Lagrangian appears in terms of the two Majorana components. Therefore chiral rotations belonging to the group $U(1)_A\otimes SU(2)$ are a symmetry at the classical level of the $N_f=1$ AdjQCD action. 

At the quantum level, the group of axial symmetry transformations leaving the partition function invariant is $Z_{2N_c}\otimes SU(2)$. This remaining symmetry can be spontaneously broken by a non-vanishing expectation value of the chiral condensate. In this case, pions emerge as massless Goldstone bosons associated to the breaking of the continuous $SU(2)$ symmetry, while the remaining discrete part implies $N_c$ degenerate coexisting manifolds of vacua. For our specific choice of the gauge group $SU(2)$ we have $N_c=2$, and the final unbroken symmetry group would be $Z_{2}\otimes SO(2)$.

\subsection{Lattice discretization}
Preserving chiral symmetry is crucial for our numerical study of the fermion condensate. However, a lattice discretization of fermion fields preserving chiral symmetry is challenging due to the limits imposed by the Nielsen-Ninomiya theorem \cite{Nielsen:1981hk}. As demonstrated in Ref.~\cite{Ginsparg-Wilson}, a modified chiral symmetry can be realized on the lattice if the continuum anticommutator of the massless Dirac operator $D=\gamma_{\mu}(\partial_{\mu}+A^{a}_\mu(x) T_a^A))$ with $\gamma_5$
\begin{equation}
    D\gamma_5 + \gamma_5 D =0\,, \label{eq:continuum chiral commutation}
\end{equation}
is modified by the addition of an irrelevant term ($a$ denotes the lattice spacing)
\begin{equation}
    D\gamma_5 + \gamma_5 D = 2a D\gamma_5D\,.
    \label{eq:Ginsparg_Wilson}
\end{equation}
The modified anticommutator, known as Ginsparg-Wilson relation, translates to a modified lattice chiral rotation of the Dirac field:
\begin{eqnarray}
    \psi'&=&\exp~(i\alpha\gamma_5( \mathbb{1}-aD))~\psi\nonumber \\ \bar{\psi'}&=&\bar{\psi}~\exp~(i\alpha(\mathbb{1}-aD)\gamma_5)\, .
    \label{eq: modified chiral rotation}
\end{eqnarray}
Using the Ginsparg-Wilson relation Eq.~\eqref{eq:Ginsparg_Wilson} one can verify that the transformations Eq.~\eqref{eq: modified chiral rotation} leave the Lagrangian Eq.~\eqref{eq: Majorana_Lagrangian} invariant. However, the chiral condensate for Ginsparg-Wilson fermions,
\begin{align}
    \Sigma\equiv\braket{\bar{\psi}(\mathbb{1}-D)\psi}\, ,
    \label{eq: chiral condensate}
\end{align}
transforms non-trivially under Eq.~\eqref{eq: modified chiral rotation} and can be considered as an order parameter for chiral symmetry breaking. 

A possible solution of the Ginsparg-Wilson relation is the massless overlap operator
\begin{equation}
    D_{ov}=\frac{1}{2}+\frac{1}{2}~\gamma_5~\text{sign}(D_{\text{H}}(\kappa))\,,
\end{equation}
defined through the Hermitian Dirac-Wilson operator $D_{\text{H}}=\gamma_5D_W(\kappa)$,
\begin{align*}
    D_W = &\mathds{1} - \kappa \big[(1-\gamma_\mu)(V_\mu(x))\delta_{x+\mu,y} + \\
    &(1+\gamma_\mu)(V^ \dagger{}_\mu(x-\mu))\delta_{x-\mu,y}\big]\,.
\end{align*}
$V_\mu(x)$ are the links in the adjoint representation and the parameter $\kappa$ of the Dirac-Wilson operator used inside the sign function is an extra parameter taking values $\kappa\in[0.125,0.25]$. It appears as a freedom in choosing the overlap operator and can be tuned to improve locality. A more practical way to write the sign function is through the inverse square root
\begin{align*}
    \text{sign}(D_{\text{H}})=\frac{D_{\text{H}}(\kappa)}{\sqrt{D_{\text{H}}(\kappa)~D_{\text{H}}(\kappa)}}\, .
\end{align*}
Unfortunately the evaluation of the square root of $D_{\text{H}}$ is computationally demanding. Furthermore, the force is ill-defined around the origin of the spectrum, which can lead to numerical problems when integrating the equation of motion for the Hybrid Monte-Carlo (HMC) algorithm \cite{Cundy:2005pi,Wenger:2006ps}.

In our simulations we implement overlap fermions using a polynomial approximation of order $N$ of the sign function, following the algorithm described in Ref.~\cite{piemonte_monte-carlo_2020}. At finite $N$ the Ginsparg-Wilson equation Eq.~\eqref{eq:Ginsparg_Wilson} is only approximately fulfilled which introduces an explicit breaking of the chiral symmetry. The quality of the polynomial approximation can be visually seen when studying the eigenvalues of the overlap operator, see Fig.~\ref{fig:eigenvalues}. An advantage of this approximation is that it introduces a gap on the spectrum, which acts as an IR regulator, preventing the forces on the HMC algorithm to diverge. As $N$ is increased, the approximation converges to the exact one and in the limit $N\rightarrow\infty$, the spectral gap disappears, the chiral point is reached and the (modified) chiral symmetry of Eq.~\eqref{eq:Ginsparg_Wilson} is restored. There are several advantages in our approach compared to standard Wilson fermions:
\begin{itemize}
\item there is no need of fine tuning of the fermion mass, as the chiral limit is reached after a simple $N\rightarrow\infty$ extrapolation,
\item as such, we can study chiral symmetry breaking directly using the chiral condensate as order parameter without having to worry about additive renormalization terms,
\item the lattice action is automatically $O(a)$ improved.
\end{itemize}
We have chosen periodic boundary conditions applied to all fields, motivated by the suppression of finite size effects observed from our previous experience in supersymmetric models. In particular, we expect a certain degree of cancellation between fermion and boson states even if supersymmetry is completely broken in our model by the infinite mass given to the scalar fields
\cite{Unsal:2008eg}.

For the discretization of the gauge part of continuum action we use a tree level Symanzik improved action. Even though the strongest source for lattice artifacts is the fermionic action, the Symanzik improvement helps to evade spurious phase transitions that could potentially appear when studying gauge observables like the Polyakov loop.

\subsection{Parameter tuning}

\begin{figure}
	\centering\includegraphics[width=0.49\textwidth]{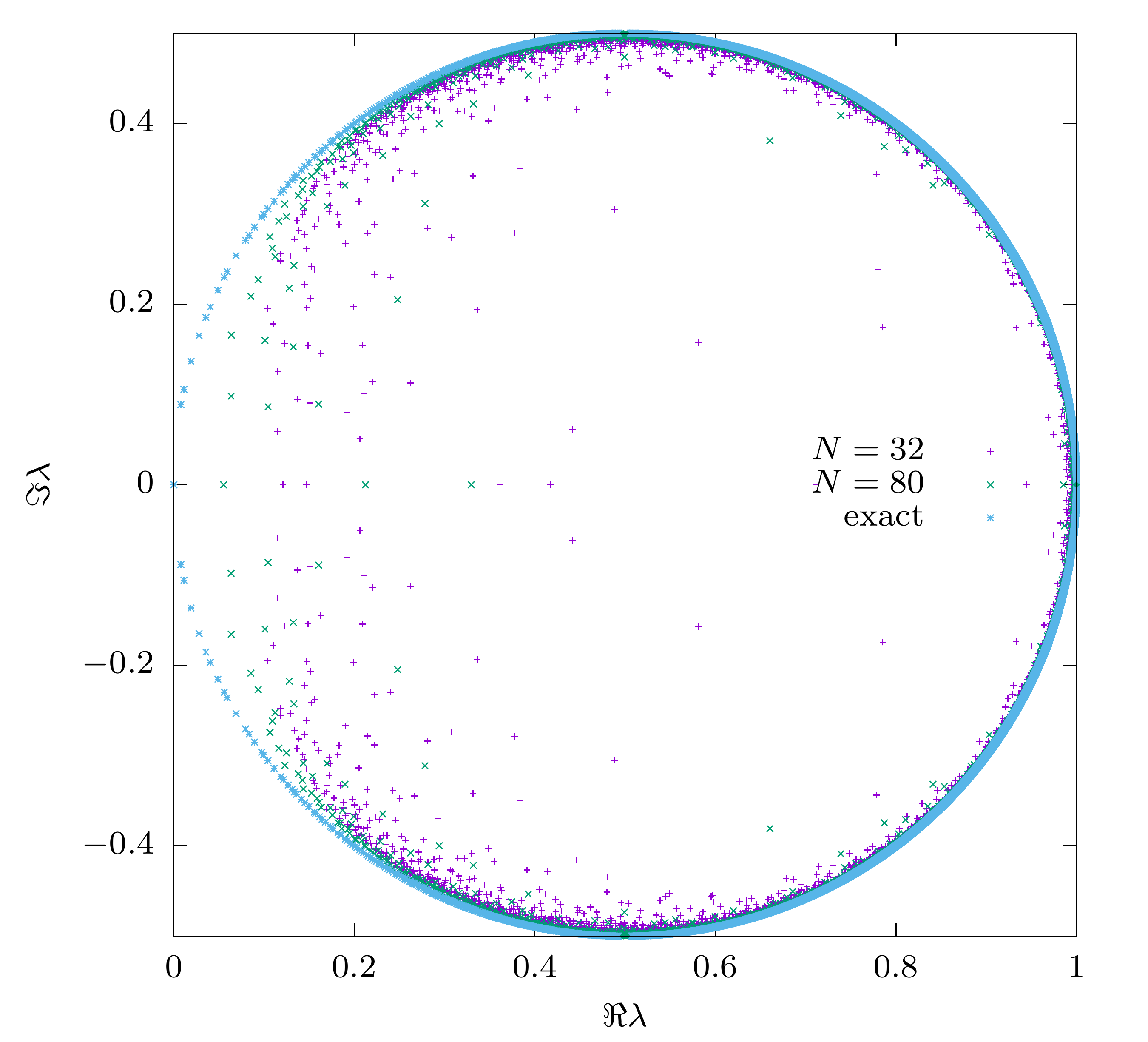}
	\caption{Spectrum of the exact and the approximated overlap operator for $N$ equal to 32 and 80 on the complex plane ($8^4$ lattice at $\beta=1.75$).}
	\label{fig:eigenvalues}
\end{figure}

The critical behavior of a renormalization group transformation of AdjQCD near the Gaussian fixed point is dominated by two relevant parameters, namely the gauge coupling and the fermion mass, if the number of flavors $N_f$ is below the critical threshold where asymptotic freedom is lost. The fermion mass is a relevant direction even in the vicinity of the infrared fixed point inside the conformal window. In the Wilsonian low energy effective action, a mass term is generated near a fixed point from the violation of the Ginsparg-Wilson relation induced by our polynomial approximation, a mass that is going to vanish in the limit $N \rightarrow \infty$.

\begin{figure}
    \centering
    \subfigure[$\beta=1.6$]{
    \includegraphics[width=0.49\textwidth]{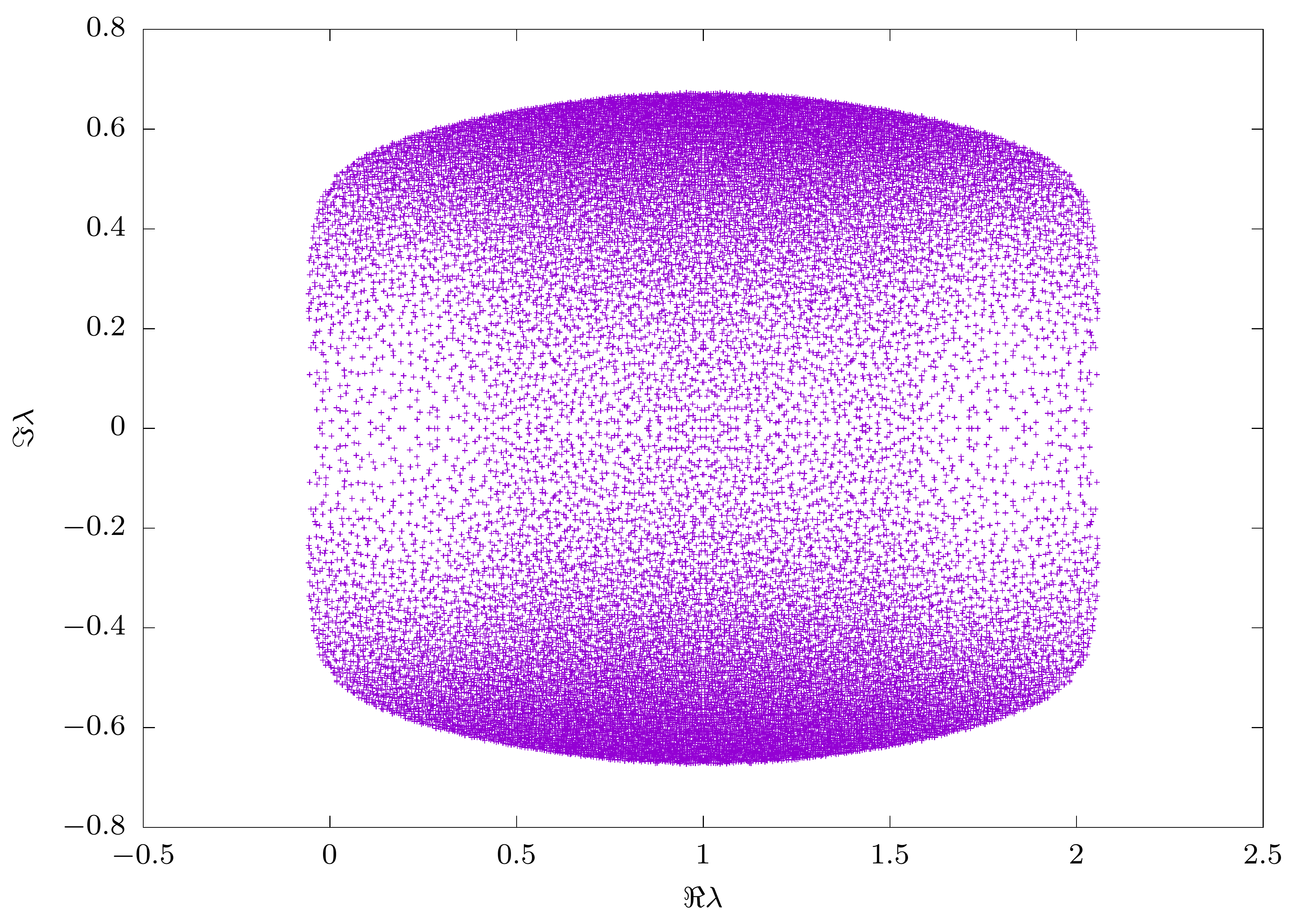}}
    \subfigure[$\beta=1.75$]{
    \includegraphics[width=0.49\textwidth]{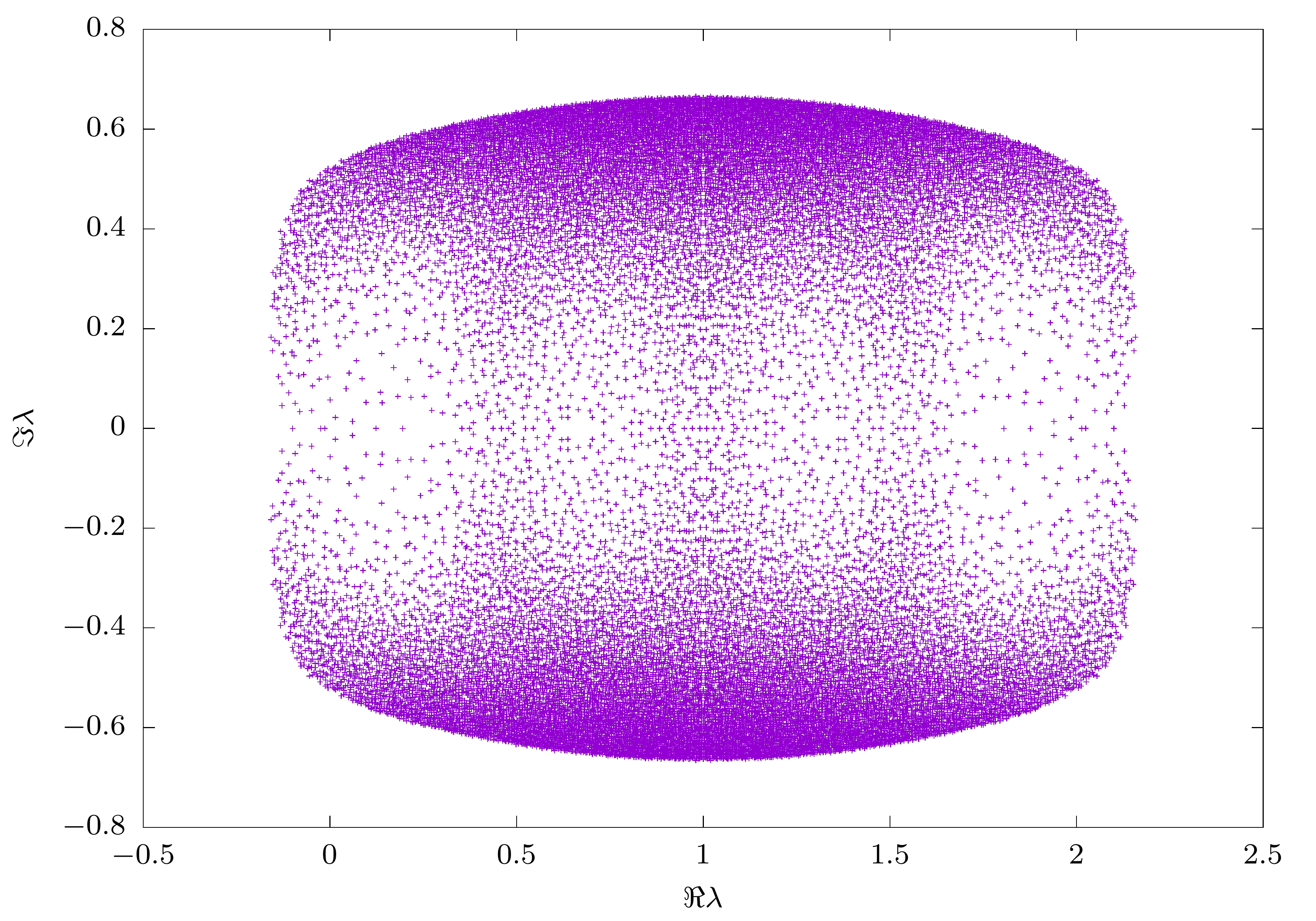}}
    \caption{Spectrum of the Dirac-Wilson operator ($D_W$) measured on a single configuration. The configurations have been generated with the approximated overlap operator ($N=80$) on a $8^4$ lattice.}\label{fig:spectrum_dirac}
\end{figure}

We have set the hopping parameter $\kappa$ of the Dirac-Wilson operator inside the sign function to 0.2, and we apply one level of stout smearing to the corresponding link in the adjoint representation with a parameter $\rho=0.15$. We have verified that the overlap operator has zero eigenvalues, and our polynomial approximation converges toward the expected circle while keeping all eigenvalues inside it, see Fig.~\ref{fig:eigenvalues}. The spectrum of the Dirac-Wilson operator is quite scattered and dense at small $\beta$, while converging toward the expected shape with four holes, one of them lying around the origin of the complex plane, see Fig.~\ref{fig:spectrum_dirac}. The projection to the unit circle leads therefore to a single Dirac fermion interacting with the gauge fields \cite{Creutz:2006ts}. 

\begin{table}
    \centering
    \begin{tabular}{|c|c|c|c|c|c|}
\hline
$N$ & $L$ & $\beta$ & $a^3 \Sigma$ & $\langle P \rangle$ & $\langle P_L \rangle$\\
\hline
250 & 4 & 1.45 & 0.00022(25) & 0.50504(57) & 0.0567(18) \\
250 & 4 & 1.5 & 0.00295(62) & 0.52223(59) & 0.0582(17) \\
250 & 4 & 1.55 & 0.0023(69) & 0.53911(62) & 0.0614(19) \\
250 & 4 & 1.6 & 0.00273(66) & 0.55582(65) & 0.0645(19) \\
250 & 4 & 1.65 & 0.006(1) & 0.57485(67) & 0.0671(24) \\
250 & 4 & 1.7 & 0.0056(22) & 0.59275(66) & 0.0781(22) \\
250 & 4 & 1.75 & 0.0053(23) & 0.61288(92) & 0.0943(37) \\
250 & 4 & 1.8 & 0.0059(17) & 0.63417(75) & 0.1269(50) \\
250 & 4 & 1.85 & 0.0012(15) & 0.65579(71) & 0.1587(55) \\
250 & 4 & 1.9 & 0.0045(41) & 0.66946(67) & 0.1811(55) \\
250 & 6 & 1.4 & 0.00024(8) & 0.48919(27) & 0.02762(73) \\
250 & 6 & 1.45 & 0.00079(8) & 0.50542(24) & 0.02858(77) \\
250 & 6 & 1.5 & 0.00237(19) & 0.52208(27) & 0.02775(69) \\
250 & 6 & 1.55 & 0.00533(52) & 0.53772(28) & 0.02832(78) \\
250 & 6 & 1.6 & 0.00866(29) & 0.55461(2) & 0.02868(56) \\
250 & 6 & 1.65 & 0.0157(11) & 0.5724(3) & 0.0305(9) \\
250 & 6 & 1.7 & 0.0214(15) & 0.59402(36) & 0.0348(13) \\
250 & 6 & 1.75 & 0.0225(15) & 0.61602(40) & 0.0325(14) \\
250 & 6 & 1.8 & 0.0118(28) & 0.6401(4) & 0.0479(18) \\
250 & 6 & 1.85 & 0.00128(41) & 0.65966(28) & 0.0724(26) \\
250 & 6 & 1.9 & 0.00068(34) & 0.67341(25) & 0.0856(35) \\
\hline
    \end{tabular}
    \caption{Chiral condensate, plaquette and Polyakov loop of the small volumes runs used for tuning the bare lattice gauge coupling.}
    \label{tab:small_volume}
\end{table}

We have also verified that all our simulations are in the confined phase in the region of bare couplings we have explored even at volumes as small as $6^4$, see Tab.~\ref{tab:small_volume}, and free from possible bulk phase transitions.

\section{Chiral symmetry breaking}\label{sec:chiral}
\subsection{The chiral condensate}

The chiral condensate is equal to the derivative of the partition function with respect to the quark mass. As we are interested in understanding whether the theory is chirally broken, an important indication is a non-zero value of the chiral condensate extrapolated to the $N\rightarrow\infty$ limit.

First, we have explored the behavior of the chiral condensate $\Sigma$ as a function of the volume. We have found that in the region of the bare couplings up to $\beta=1.75$ there is no clear evidence of finite volume effects up to a lattice of size $8^4$, while at $\beta=1.8$ larger lattice sizes are required. On all runs used for the extrapolation of the chiral condensate to the massless limit, the Polyakov loop expectation value is zero and the theory is therefore in the confined phase.

\begin{figure*}
    \centering
    \subfigure[$\beta=1.6$]{
    \includegraphics[width=0.47\textwidth]{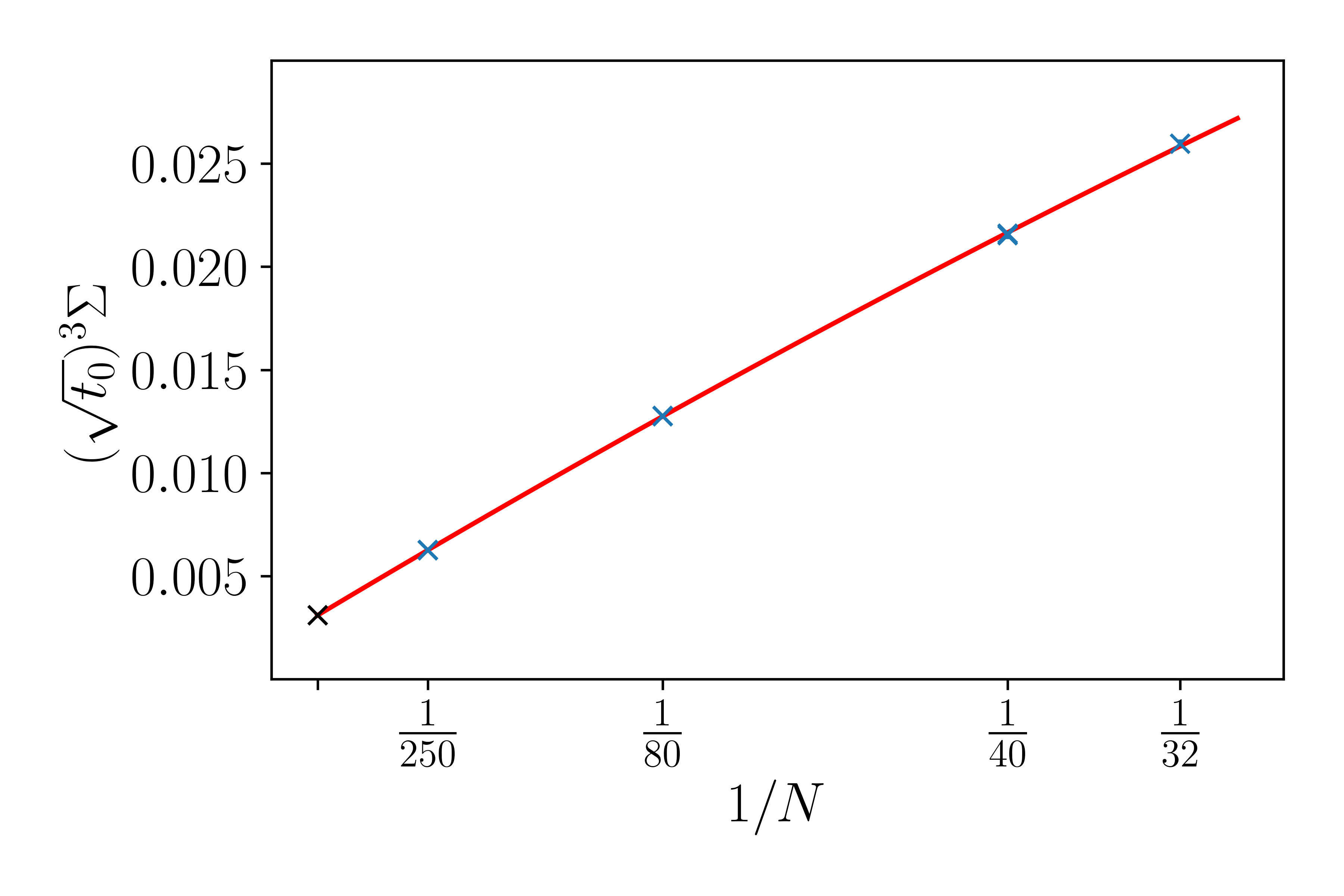}}
    \subfigure[$\beta=1.7$]{
    \includegraphics[width=0.47\textwidth]{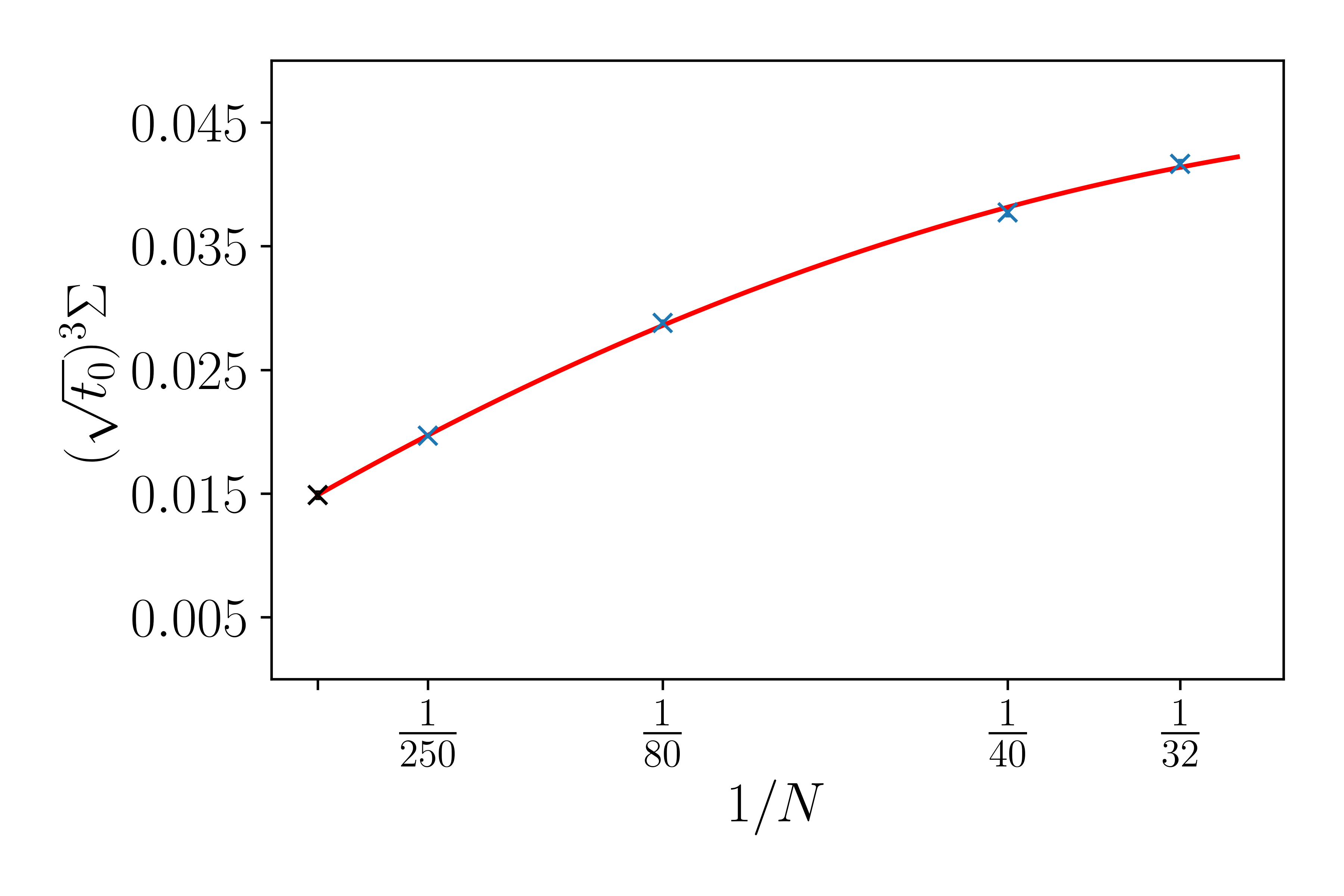}}
    \subfigure[$\beta=1.75$]{
    \includegraphics[width=0.47\textwidth]{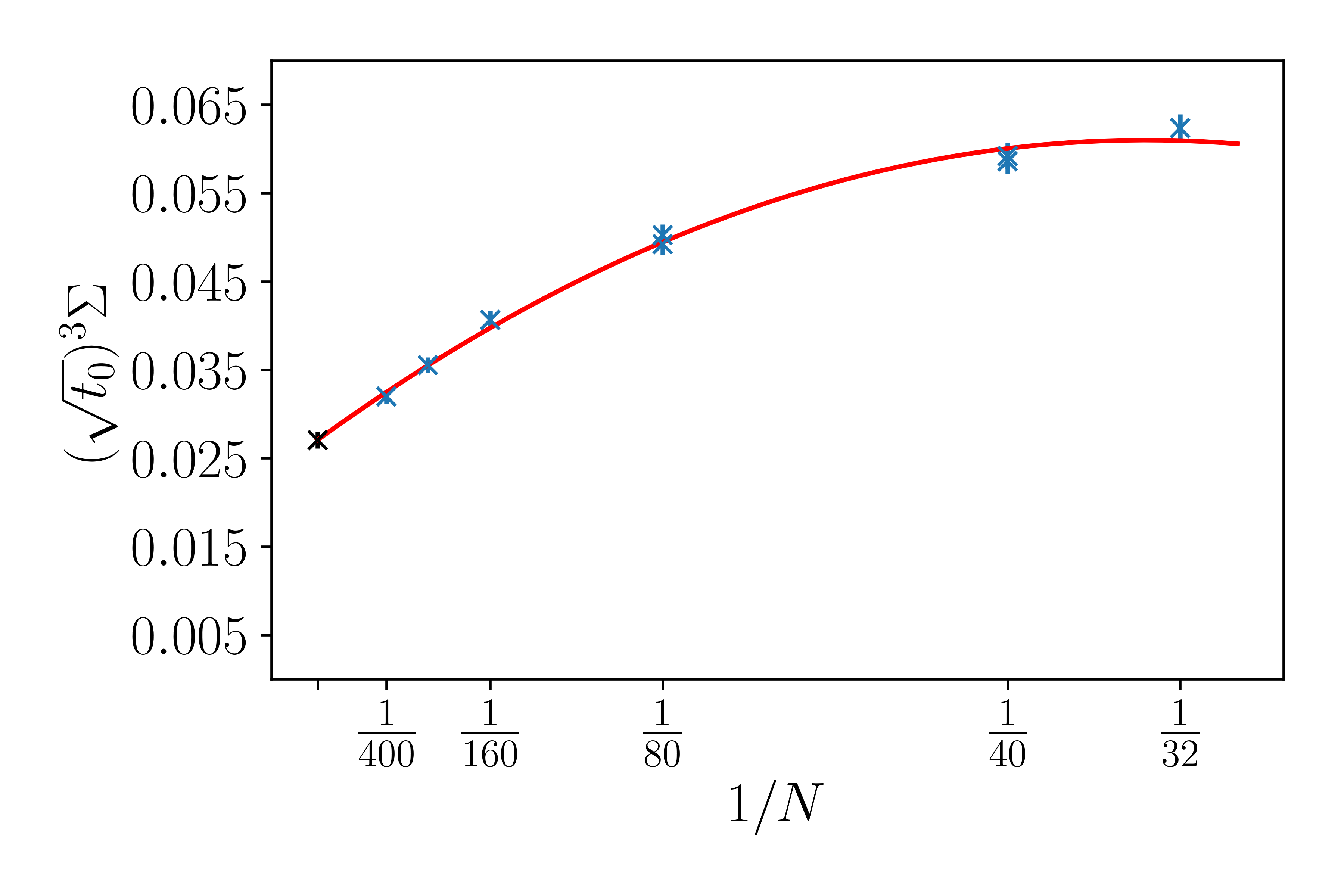}}
    \subfigure[$\beta=1.8$]{
    \includegraphics[width=0.47\textwidth]{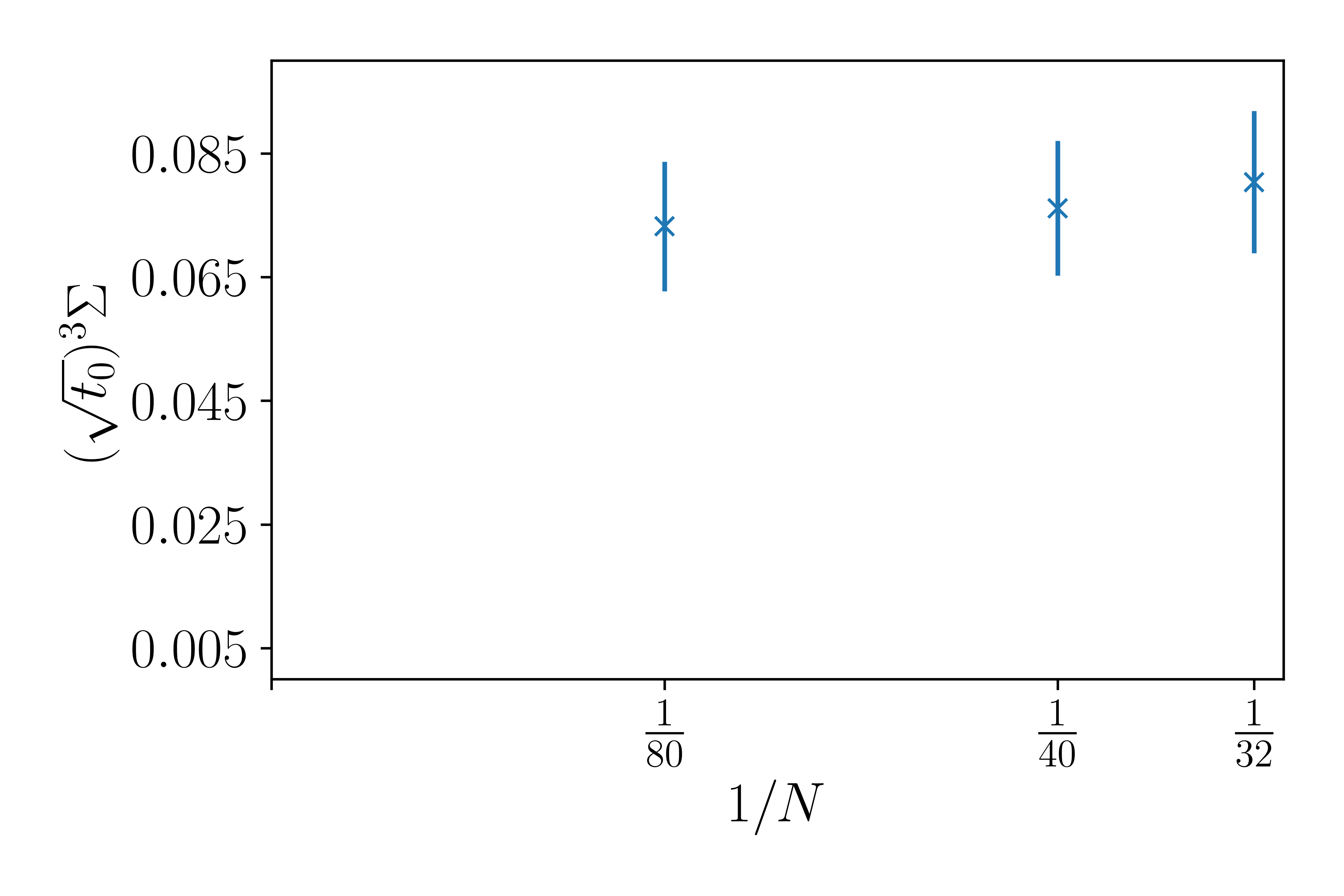}}
    \caption{Extrapolation of the bare chiral condensate in dimensionless units to the chiral limit, including in the fit the volumes $V=8^4$ and $V=18^4$. The value of the scale $\sqrt{t_0}/a$ is extrapolated first to the chiral limit, see Sec.~\ref{sec:t0} for further details.}
    \label{fig:condensate_chiral_limit}
\end{figure*}

We have extrapolated the bare chiral condensate as a function of $1/N$ to $N=\infty$ limit at three different $\beta$. We have fitted a quadratic function with a $\chi^2/$d.o.f.\ smaller than two. The extrapolated condensate is non-zero indicating that at zero temperature chiral symmetry is spontaneously broken, see Fig.~\ref{fig:condensate_chiral_limit}. As we have observed in the case of $\mathcal{N}=1$ Super-Yang-Mills, the main contribution for the non-vanishing expectation value of $\Sigma$ comes from a smaller and smaller number of configurations as one approaches $N\rightarrow\infty$. This pattern means that we have been able to effectively regularize the ``zero over zero'' problem of simulating massless fermions using a polynomial approximation.

We plan to complete our simulations at $\beta=1.8$ and to extend them at $\beta=1.85$, in order to extrapolate the renormalized value of the chiral condensate to the continuum limit in future studies. Here we can note that the bare value of $\Sigma$ in dimensionless units is increasing as a function of the bare gauge coupling. Assuming that chiral symmetry breaking persists in the continuum limit, this changing of $\langle\bar{\psi}\psi\rangle$ implies a large anomalous dimension, as it has been observed in previous investigations with Dirac-Wilson operator \cite{Georg:2015}.

\subsection{Chiral rotations}

As shown in the previous section, chiral symmetry is spontaneously broken by a non-vanishing vacuum expectation value of the chiral condensate in our simulations. The remaining symmetry corresponds to $Z(2)\otimes SO(2)$. The discrete part $Z(2)$ corresponds to two manifolds of vacuum states distinguished by a positive and negative values of the chiral condensate. It is quite challenging to observe the coexistence of these phases directly in our simulations, as we are approaching the chiral limit from small but positive effective fermion masses. Nevertheless, if we perform a chiral rotation by an angle $\pi$, we expect to map in the limit $N\rightarrow\infty$ a configuration from positive to negative value of the chiral condensate, leaving its absolute value invariant. We can use the deviation at fixed $N$ of the absolute value of the chiral rotated condensate as a measure of how much chiral symmetry is broken by the polynomial approximation of the sign function. Alternatively, we can view the rotation of the chiral condensate as a chiral Ward identity where the angle is chosen in order to avoid to take into account the effects of axial anomaly.

\begin{figure}[t]
    \centering
    \subfigure[$\beta=1.6$]{\includegraphics[width=0.47\textwidth]{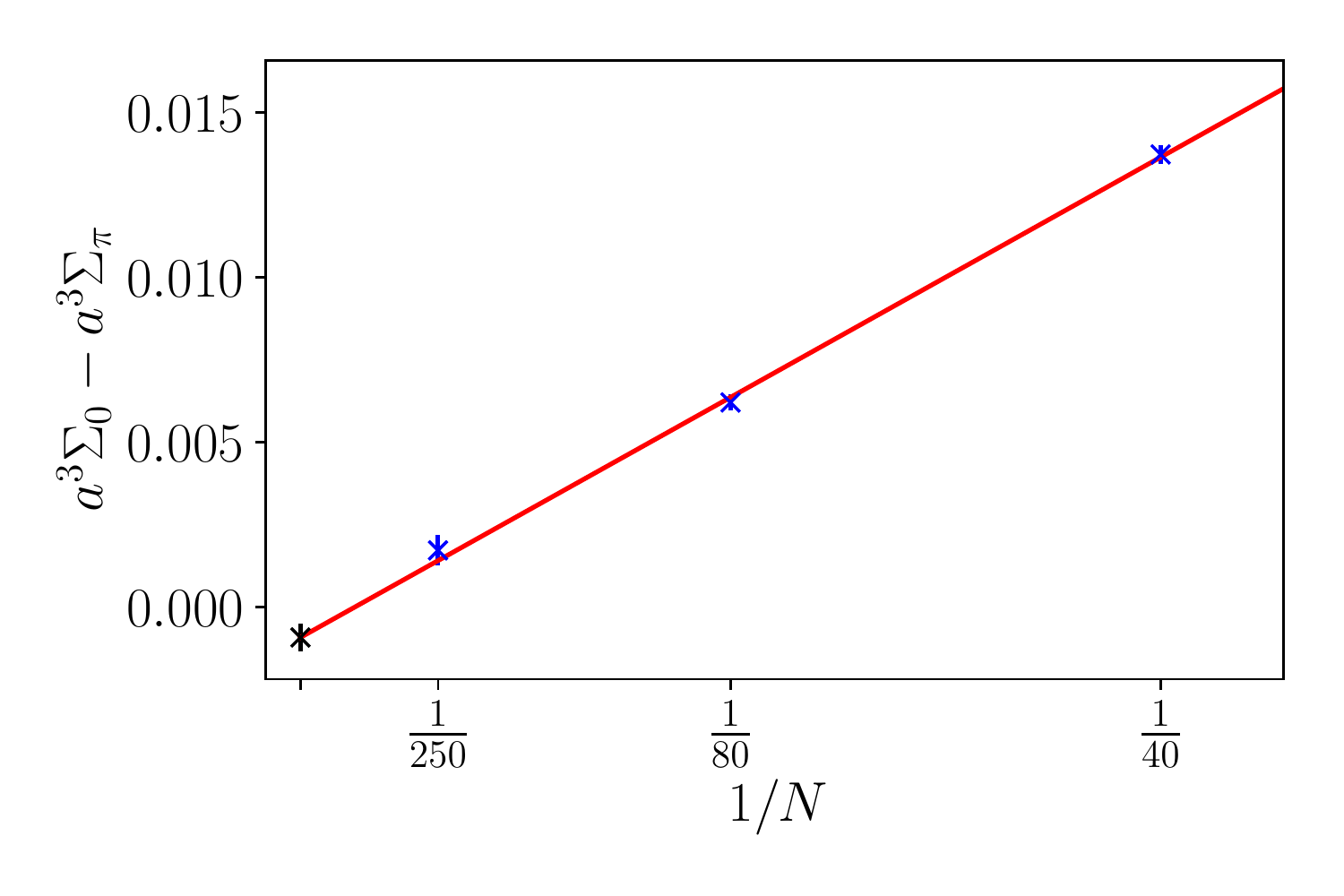}}
    \subfigure[$\beta=1.7$]{\includegraphics[width=0.47\textwidth]{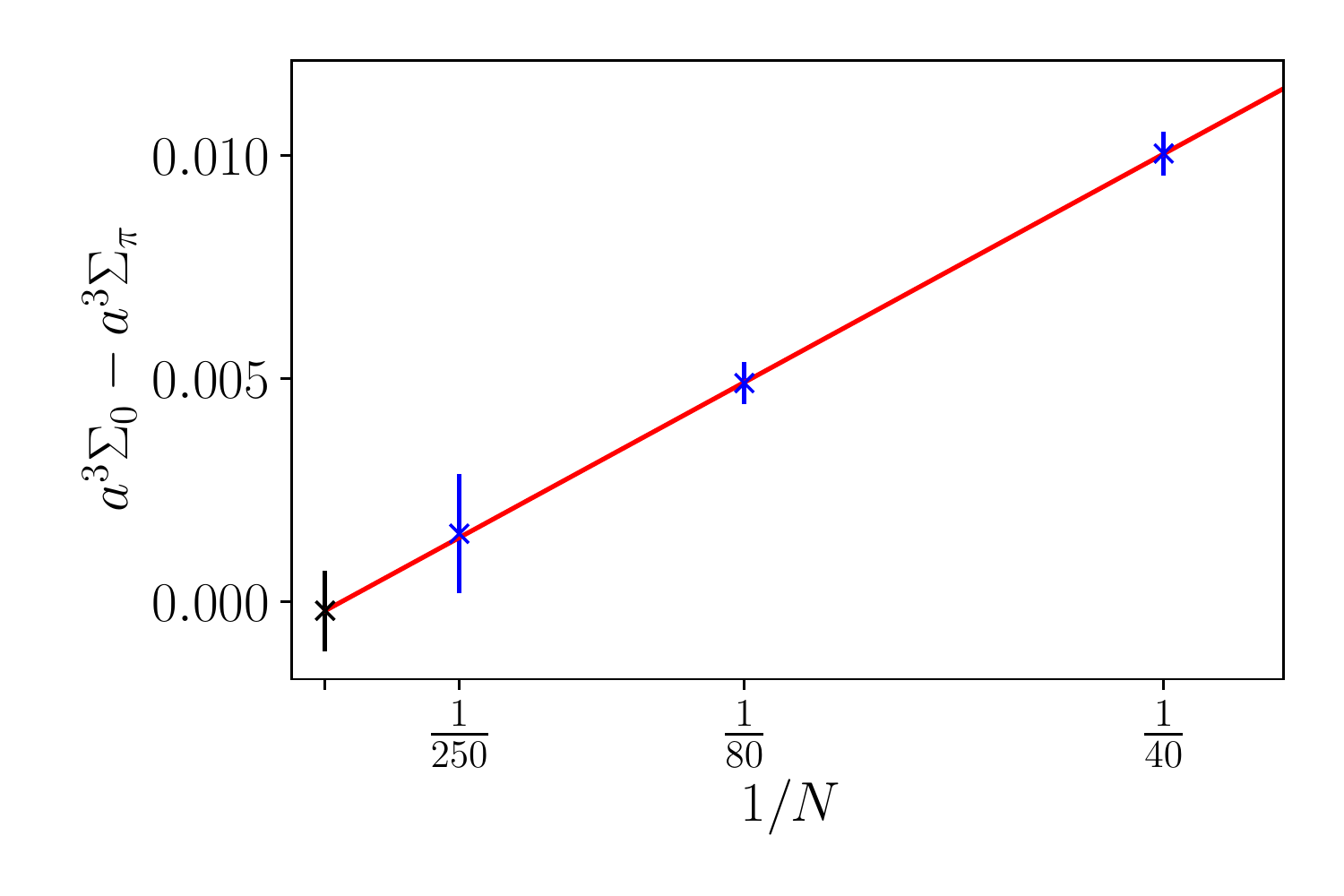}}
\caption{Sum of the bare fermion condensate $\Sigma$ in lattice units before and after a chiral rotation by an angle $\omega=\pi$. The final extrapolated value of the sum is equal to -0.0009(4) at $\beta=1.6$ and -0.0002(9) at $\beta=1.7$.}
\label{fig:chiral_rotation}
\end{figure}

In order to perform a chiral rotation numerically, we approximate Eq.~\eqref{eq: modified chiral rotation} for a small chiral rotation as
\begin{eqnarray}
    \psi & \rightarrow & \psi'=\psi  + i\omega\gamma_5(\mathbb{1} -a D))\psi\,, \nonumber\\
    \bar{\psi} & \rightarrow & \bar{\psi'}=\bar{\psi} + i\omega \bar{\psi}((\mathbb{1} -a D)
    \gamma_5)\,,
    \label{eq: finite chiral rotation}
\end{eqnarray}
such that we can decompose a chiral rotation of an angle $\alpha$ into $n$ small steps $\alpha= \sum_n\omega $ by applying repeatedly Eq.~\eqref{eq: finite chiral rotation} as many times as needed. We have computed the chiral condensate $\Sigma_0$ for a random source $\psi$ and a rotated chiral condensate $\Sigma_{\pi}$ using the same sources rotated by an angle $\alpha=\pi$. The difference in absolute value between the two condensates $\Delta\Sigma=|\Sigma_0|- |\Sigma_{\pi}| $ is shown in Fig.~\ref{fig:chiral_rotation}.

As we can see, the combination $\Delta\Sigma$ gives a non-zero value for a finite $N$. As we increase the order of the polynomial approximation, the distance to the chiral point decreases. In the limit $N\rightarrow\infty$ the chiral condensate changes by a sign flip and $\Delta\Sigma$ extrapolates to a value compatible with zero, pointing out that the $Z_2$ symmetry is recovered and the chiral symmetry has been restored.

\section{Scale setting and running of the strong coupling constant}\label{sec:flow}

We can show further evidence on the absence of a conformal behaviour if we consider the running of the strong coupling constant in the infrared limit. The strong coupling constant $\alpha_s$ is a scheme dependent quantity depending on the energy scale $\mu$. Perturbation theory is recovered near the Gaussian fixed point as $\mu\rightarrow\infty$, while the infrared behaviour in the limit $\mu\rightarrow0$ is dominated by non-perturbative effects. If the theory is infrared conformal, we expect to see a freezing of the running of $\alpha_s(\mu)$ as the energy scale approaches zero. We want to verify from our lattice simulations whether this scenario is realized. We have investigated the running of the strong coupling directly, and we have also tried to determine whether the theory has an infrared fixed point by extrapolating the lattice scale to the chiral limit and observing whether it vanishes or not. In the following two subsections we consider the scale setting defined from the gradient flow, which enables us at the same time to measure also the running of the strong coupling constant.

\subsection{Scale setting}\label{sec:t0}

\begin{figure}
    \centering
    \includegraphics[width=0.47\textwidth]{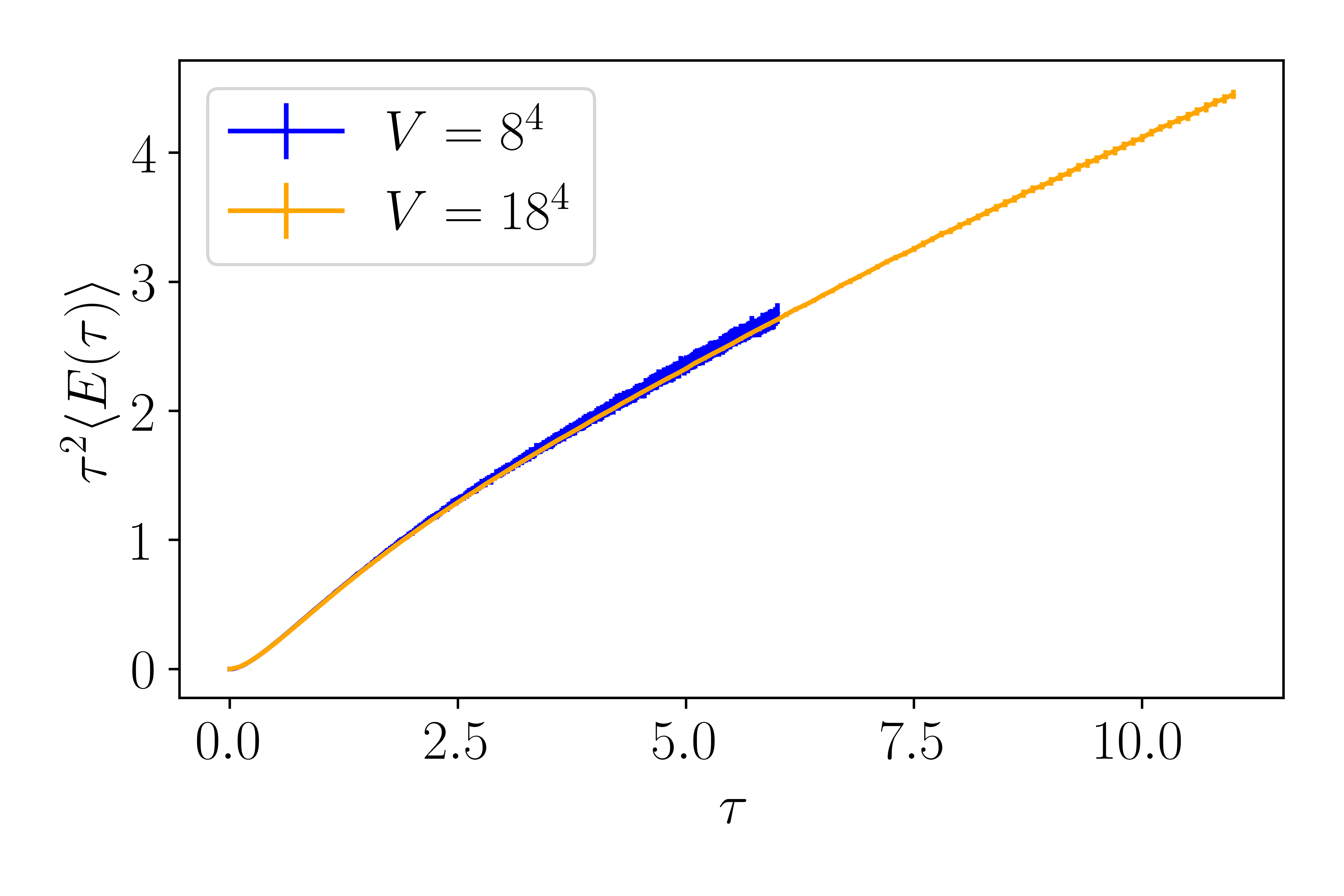}
    \caption{Comparison of the flowed energy density for the volume $V=18^8$ and $8^8$ with $N=40$ at $\beta=1.6$.}
    \label{fig:flow_finite_volume_comparison}
\end{figure}

The gradient flow is a continuous smoothing applied to gauge fields, defined from a partial differential equation which is solved numerically to determine the evolution of certain gauge-invariant observables as a function of the flow-time $\tau$ \cite{Luscher:2009eq}. In particular, the measure of the flowed gauge energy density
\begin{align}
    \braket{E(\tau)}=\frac{1}{4}G^a_{\mu\nu}G^a_{\mu\nu}(\tau)\,,
\end{align}
allows to define a scale $t_0/a^2$ as the flow time $\tau$ where $\tau^2\braket{E(\tau)}$ reaches the reference value 0.3. The lattice spacing is then proportional to $1/\sqrt{t_0}$, and its value in physical units, such as femtometers, could be defined in principle in terms of an experimentally measurable quantity. 

\begin{table}
\centering
\begin{tabular}{|c|c|c|c|c|}
\hline
$N$ & $L$ & $\beta$ & $a^3 \Sigma$ & $\sqrt{t_0}/a$\\
\hline
250 & 12 & 1.6 & 0.01209(15) & - \\
250 & 8 & 1.6 & 0.0121(5) & 0.8040(15) \\
250 & 8 & 1.7 & 0.02787(85) & 0.8921(19) \\
250 & 8 & 1.75 & 0.032(13) & - \\
160 & 8 & 1.75 & 0.03661(89) & 1.033(15) \\
80 & 8 & 1.6 & 0.02469(31) & 0.8069(11) \\
80 & 8 & 1.7 & 0.04078(54) & 0.9049(18) \\
80 & 8 & 1.75 & 0.0443(7) & - \\
80 & 8 & 1.8 & 0.0297(31) & - \\
40 & 8 & 1.6 & 0.04161(23) & 0.8143(19) \\
40 & 8 & 1.7 & 0.05343(25) & 0.9197(31) \\
40 & 8 & 1.75 & 0.0528(6) & - \\
40 & 8 & 1.8 & 0.038(2) & - \\
32 & 18 & 1.6 & 0.050193(3) & 0.8226(2)\\
32 & 18 & 1.7 & 0.058982(4) & - \\
32 & 18 & 1.75 & 0.056182(7)& 1.0923(14) \\
32 & 18 & 1.8 & 0.04372(17) & 1.4020(79) \\
80 & 18 & 1.6 & 0.024649(4) & - \\
80 & 18 & 1.75 & 0.04524(13)& 1.0401(27) \\
80 & 18 & 1.8 & 0.03983(22) & 1.369(12) \\
40 & 18 & 1.6 & 0.041765(3) & - \\
40 & 18 & 1.75 & 0.053316(6)& 1.0668(23) \\
40 & 18 & 1.8 & 0.04143(14) & 1.4152(71) \\
400 & 8 & 1.75 & 0.0288(16) & 1.029(9) \\
\hline
\end{tabular}
    \caption{Chiral condensate of all large volume simulations, with in addition the scale $\sqrt{t_0}/a$ for the ensembles included in the final extrapolation to the chiral and continuum limit of the running of the strong coupling.}
    \label{tab:scale_and_condensate}
\end{table}

\begin{figure}
    \centering
    \subfigure[$\beta=1.6$]{
    \includegraphics[width=0.47\textwidth]{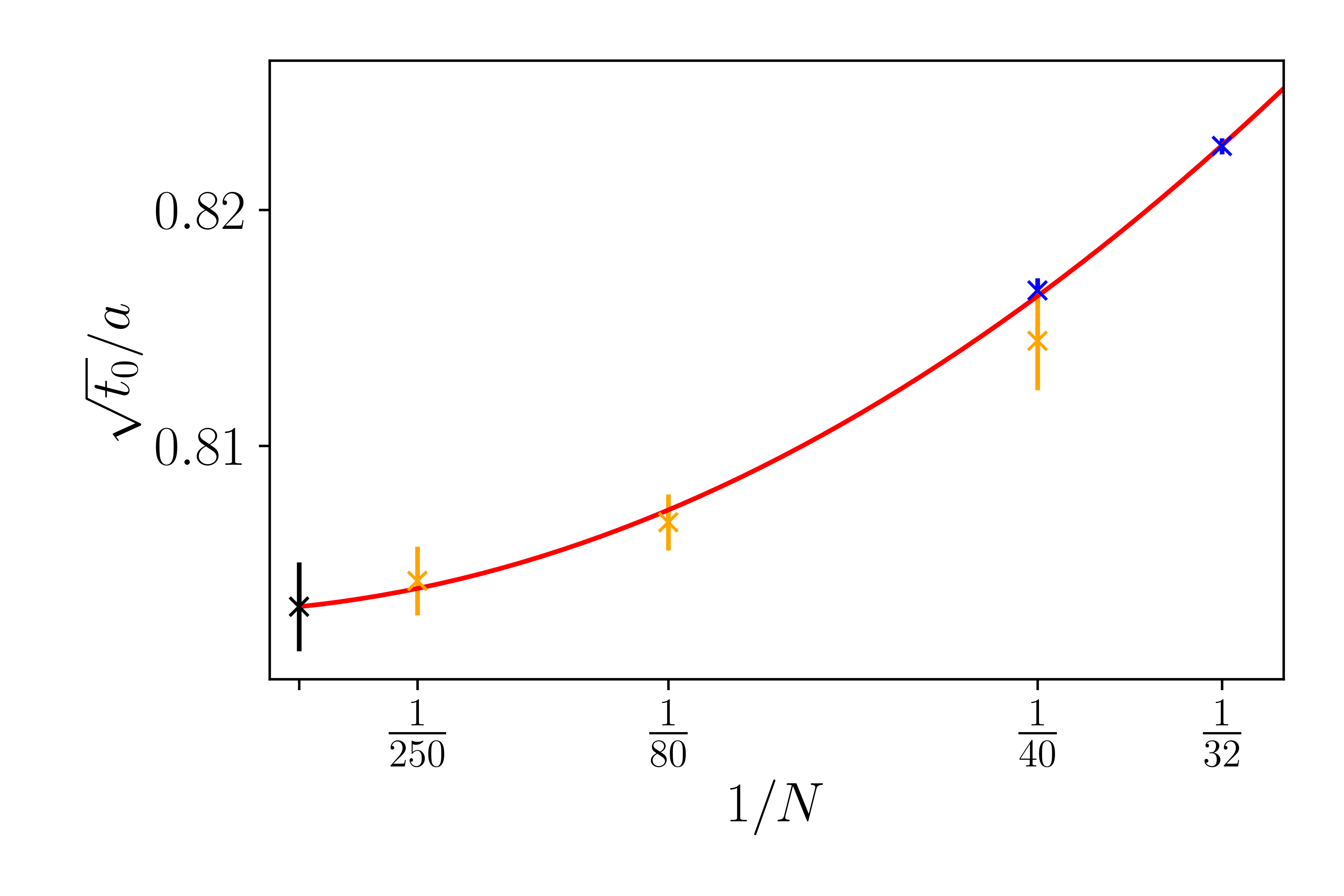}}
    \subfigure[$\beta=1.75$]{
    \includegraphics[width=0.47\textwidth]{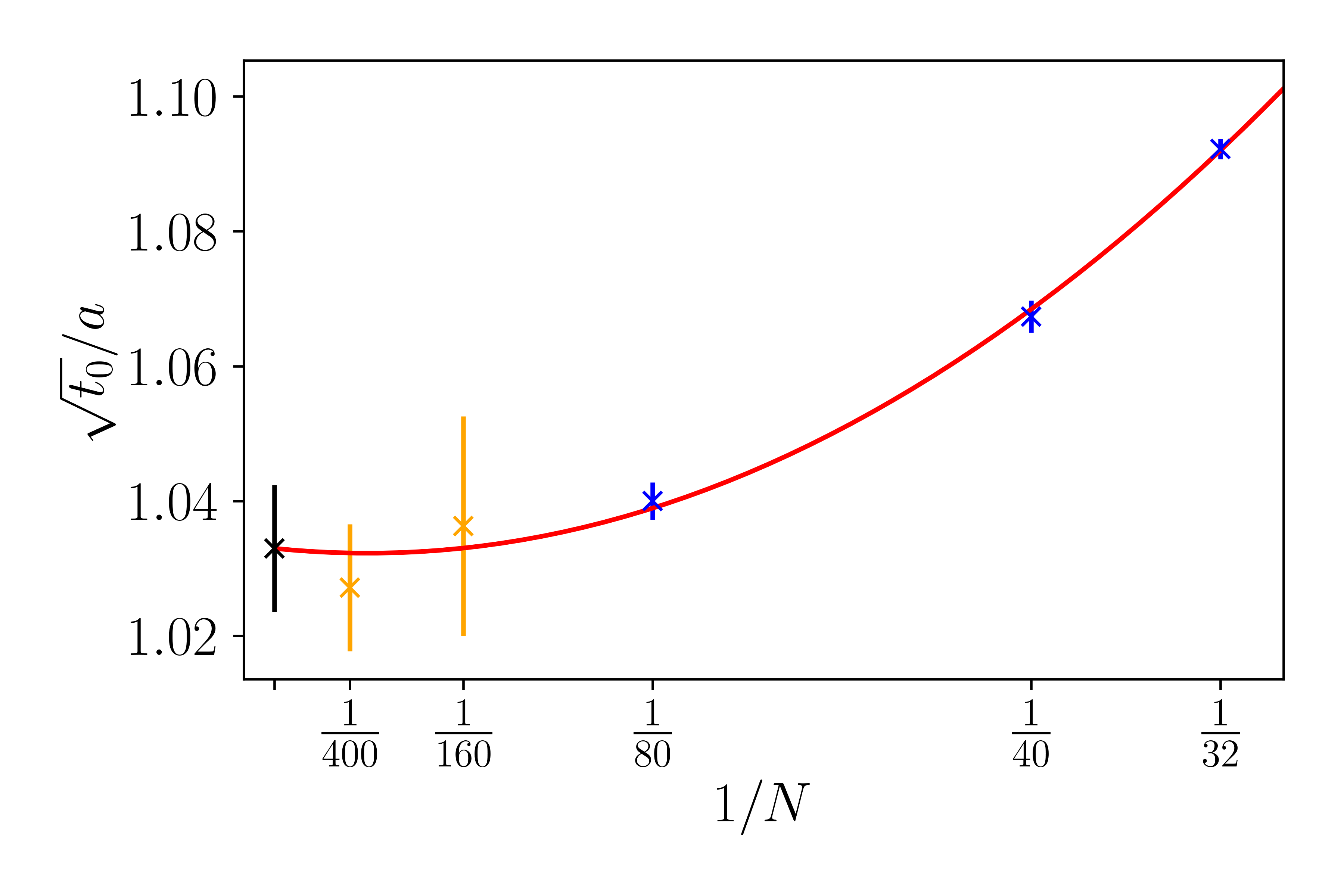}}
    \caption{Extrapolation to the chiral limit of the scale $t_0$ including in the fit the volumes $8^4$ (orange points) and $18^4$ (blue points). The errors on the larger lattices are smaller due to volume averaging, and the scales measured from both volumes are consistent.}
    \label{fig:scale_chiral_limit}
\end{figure}

We measured the scale $t_0$ using the clover plaquette discretization of the energy density and the Wilson action for the definition of the flow equations. After ensuring that finite size effects are under control, see Fig.~\ref{fig:flow_finite_volume_comparison} the extrapolation of $t_0/a^2$ to the limit $N \rightarrow \infty$ is already an indication whether $N_f=1$ AdjQCD is infrared conformal or not. In the first case, the theory does not possess any scale other than the fermion mass, and the value of $t_0/a^2$ must be zero in the chiral limit.

\begin{figure}
    \centering
    \includegraphics[width=0.47\textwidth]{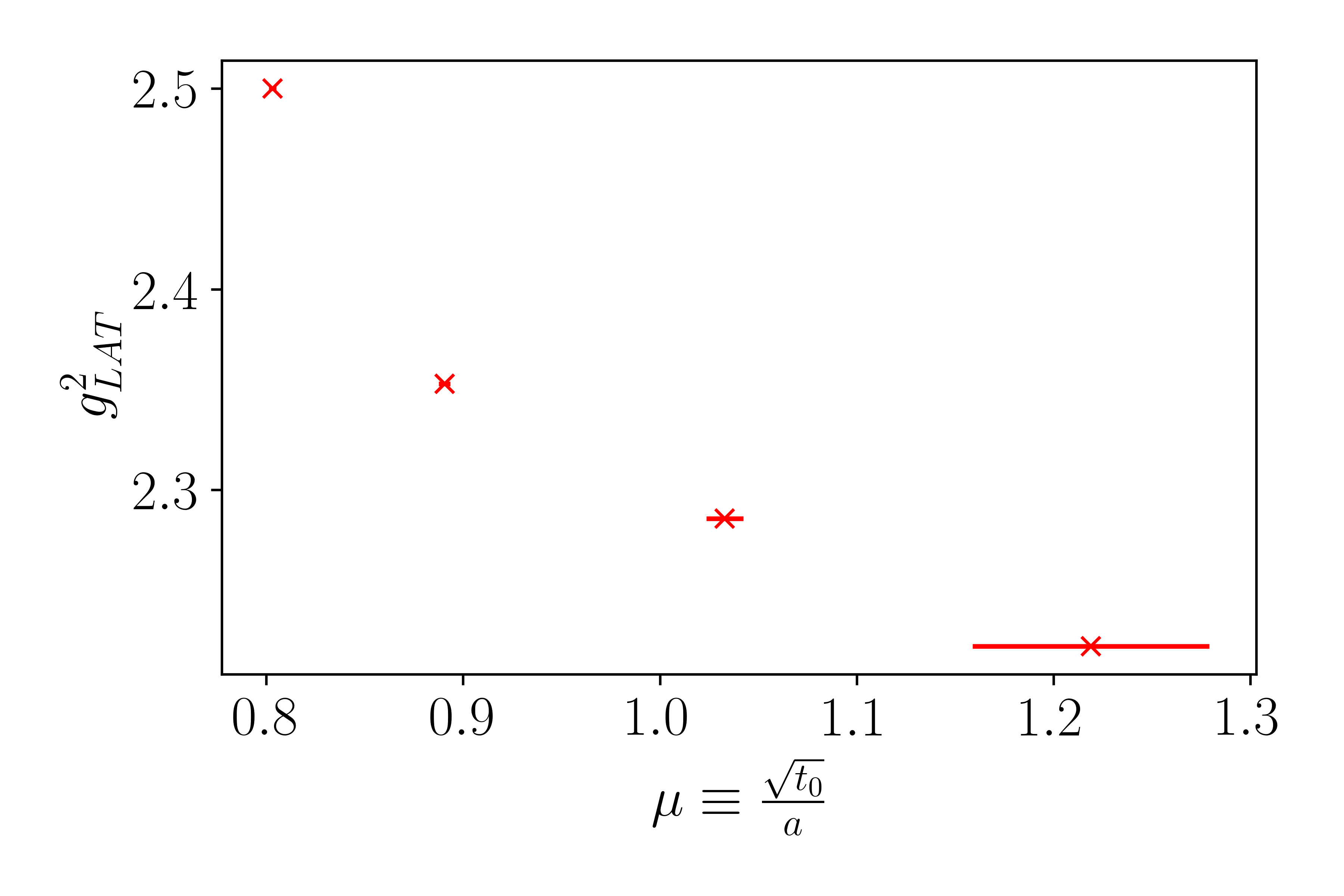}
    
    \caption{Running of the bare lattice coupling squared as a function of the inverse lattice spacing measured from the scale $\sqrt{t_0}/a$.}
    \label{fig:bare_running_coupling}
\end{figure}

The scale measured from our ensemble can be fitted by a quadratic function of $1/N$. The scale data included in the fits are summarized in Tab.~\ref{tab:scale_and_condensate} and two examples of our fits are plotted in Fig.~\ref{fig:scale_chiral_limit}. The extrapolated value of $\sqrt{t_0}/a$ is clearly different from zero, providing a first evidence that the theory is not infrared conformal. We see that $\sqrt{t_0}/a$ grows as $\beta$ is increasing, i.e.\ the lattice spacing is decreasing in the weak coupling limit as expected for a confining theory. Indeed, in the simplest way to compute the strong coupling constant scale dependence on the lattice, $\alpha_s$ is defined in terms of the bare lattice gauge coupling as $\alpha_s = \frac{2N_c}{4 \pi \beta}$, and the scale $\mu$ is equal to $1/a$. As shown in Fig.~\ref{fig:bare_running_coupling}, there are no evidence of an infrared fixed point in the region of bare coupling we have explored. However, a measure of the running of $\alpha_s$ in the low energy regime from the bare lattice coupling requires to perform simulations in the strong coupling region where possible lattice phases might prevent us from observing the true nature of the infrared fixed point. Further, this definition of the running coupling does depend on the lattice discretization of the continuum action. Fortunately, we can provide a further and cleaner evidence of the absence of an infrared fixed point by just exploiting the full dependence of the flowed gauge energy density on the flow time $\tau$.

\subsection{Running of the strong coupling constant}

\begin{figure}
    \centering
    \subfigure[$\sqrt{t_0}\mu = 0.5$]{
    \includegraphics[width=0.47\textwidth]{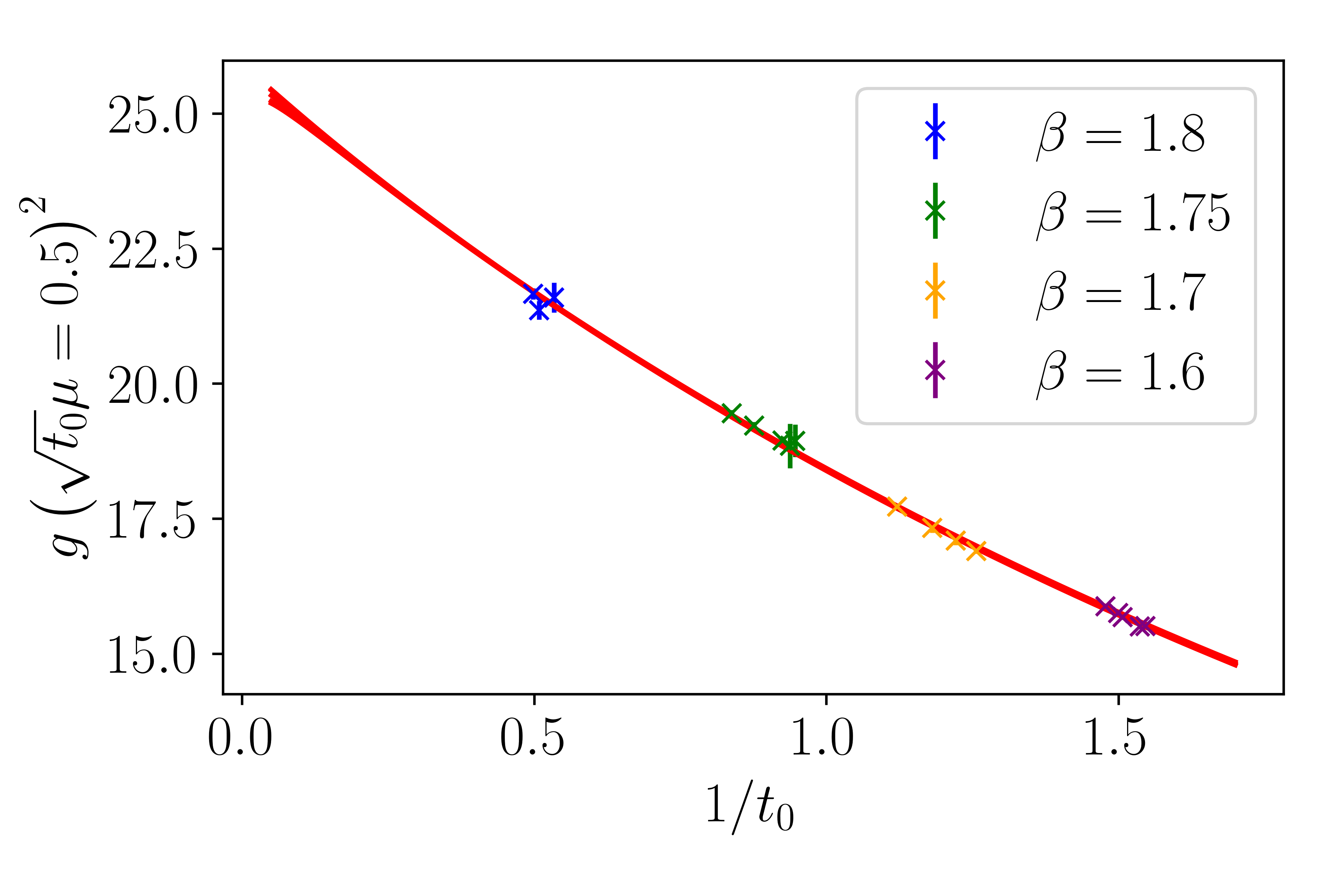}}
    \subfigure[$\sqrt{t_0}\mu = 0.25$]{
    \includegraphics[width=0.47\textwidth]{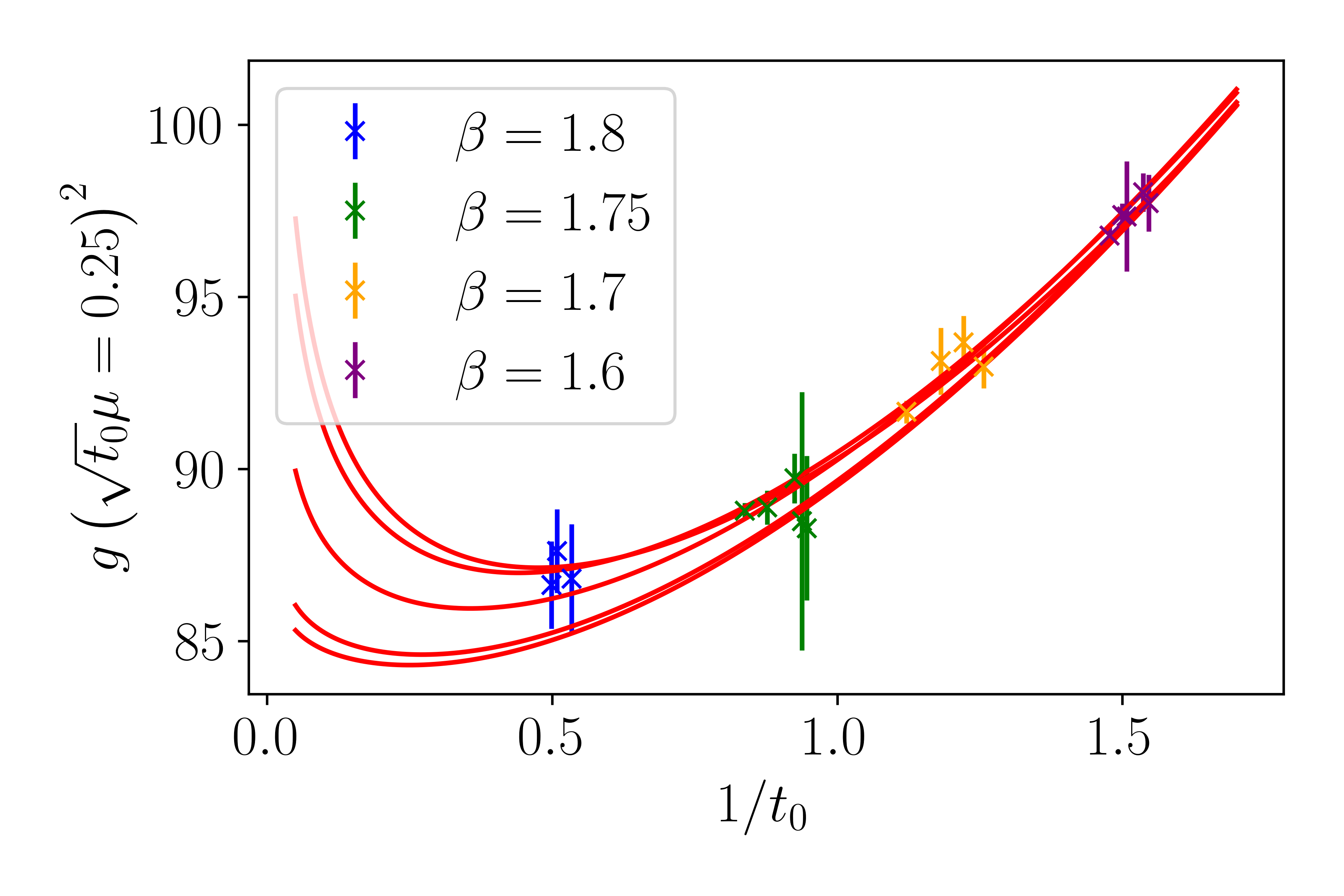}}
    \caption{Extrapolation to the chiral and continuum limit of the strong coupling constant. The red lines are obtained from the global fit by fixing the order of the polynomial approximation $N$ to 32, 40, 80, 160, and 250, from the top to the bottom. For fixed $N$, the renormalized fermion mass diverges in the continuum limit, and the coupling reaches its pure gauge value.}
    \label{fig:continuum_chiral_limit}
\end{figure}

An efficient method to compute the running of the strong coupling constant from Monte-Carlo simulations is to find an appropriate scheme which can be defined both on the lattice and in the continuum, in such a way that different computation methods and lattice discretizations can be easily compared among each other. The gradient flow can be consistently defined independently from the regularization used, and a perturbative expansion of the flowed gauge energy  density in the $\MSbar{}$ scheme reads \cite{Luscher:2009eq}
\begin{align}
     \braket{E(\tau)}=\frac{3(N^2-1)}{16\tau^2\pi^2}g^2_{\MSbar{}}(\mu)\Big(1+c_1g^2_{\MSbar{}}(\mu) + \mathcal{O}(g^4_{\MSbar{}}) \Big)\,,
\end{align}
where we have set the scale $\mu$ to be equal to $1/\sqrt{8\tau}$ \cite{Luscher:2009eq,Luscher:2010iy,Luscher:2011bx}.
This relation can be truncated to the lowest order and inverted to define a renormalized gauge coupling
\begin{align}
     g^2_{\textrm{GF}}(\mu)=\evalat*{\frac{16\pi^2}{3(N^2-1)}\tau^2\braket{E(\tau)}}{\tau^2=1/8\mu}\,.
\end{align}
The gradient flow scheme we employ requires first to extrapolate the coupling to the infinite volume limit, then to the chiral limit and finally to the continuum limit $a \rightarrow 0$ \cite{Hasenfratz:2019hpg}. This non-perturbative determination of the running of strong coupling does not depend on the lattice discretization of the continuum action we have chosen, while only the first two coefficients of the $\beta$-function are universal and scheme independent.

\begin{figure}
    \centering
    \includegraphics[width=0.47\textwidth]{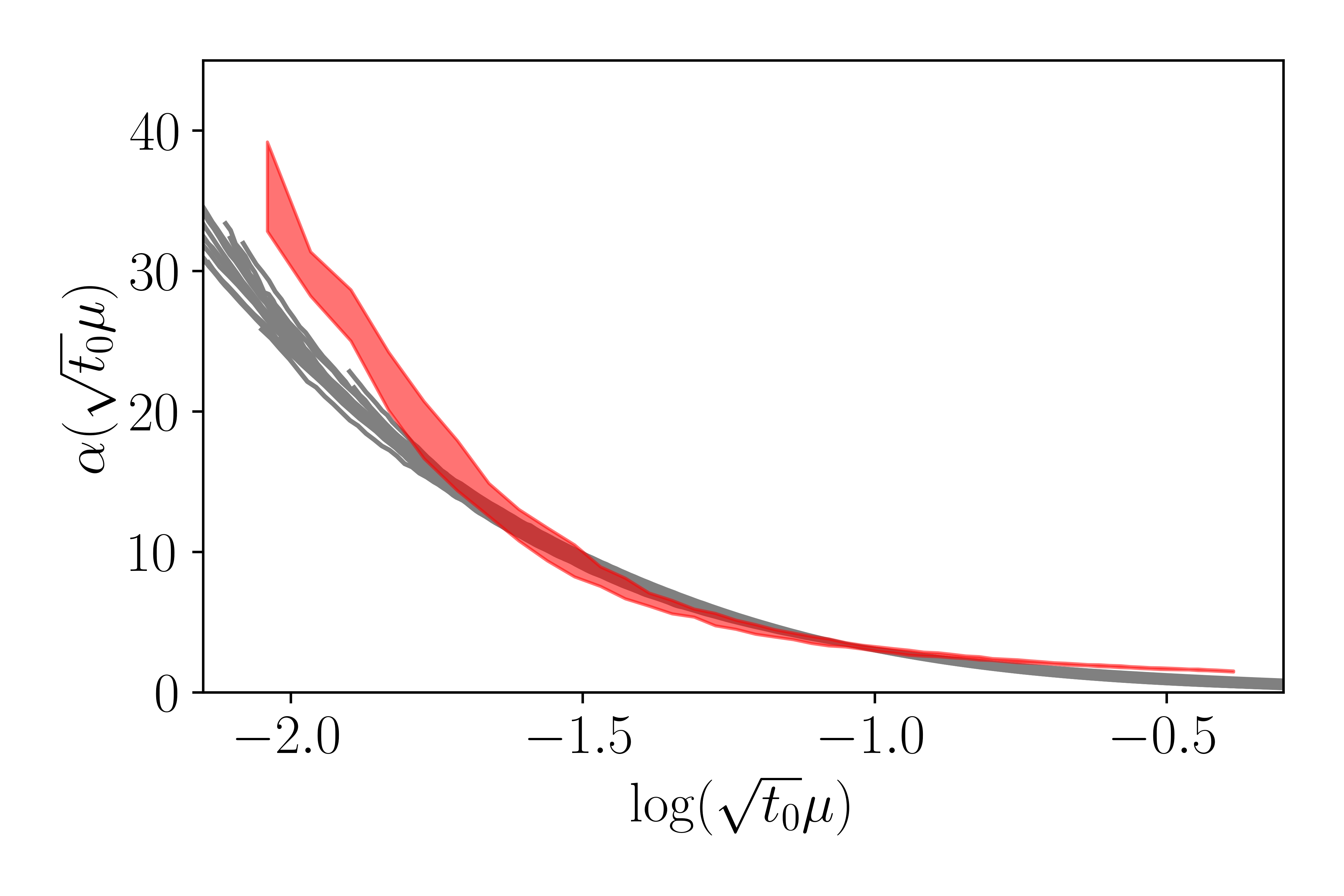}
    
    \caption{Extrapolation to the continuum limit of the strong coupling constant as a function of the scale (red band), extrapolated from the measured curves at the different $\beta$ and different polynomial approximation of the sign function (grey lines).}
    \label{fig:continuum_running_coupling}
\end{figure}

The scale $t_0$ is computed for each ensemble separately. Then, the running coupling $g^2_{GF}(\mu)$ is extrapolated to the chiral and continuum limit. We have considered a combined fit including all available ensembles in the scaling region using an ansatz of the form
\begin{equation}
g(\mu)^2 = g_0 + c_0 t_0^{-1} + c_1 t_0^{-\frac{3}{2}} + d_1 \frac{\sqrt{t_0}}{N} + d_2 \frac{1}{N} + d_3 \left(\frac{\sqrt{t_0}}{N}\right)^2\,.
\end{equation}
The fit is performed for each scale $\mu$ independently, interpolating the flow and its error in order to be able to include all ensembles at different lattice spacings in a single fit. The first term $g_0$ represents the continuum limit value of the square of the running coupling extrapolated in the limit $N\rightarrow \infty$. The terms $c_0$ and $c_1$ are lattice artefact corrections proportional to $a^2$ and $a^3$, respectively. Finally, terms corresponding to the coefficient $d_i$ are corrections proportional to a mass term, to its square and to a lattice discretization error equal to the bare fermion mass itself. At high energy, lattice artefacts are dominant, while at low energy the terms proportional to $N$ becomes relevant, see Fig.~\ref{fig:continuum_chiral_limit}. The inclusion of a finite-volume correction term, or of a logarithmic correction in the lattice spacing to the leading asymptotic scaling \cite{Husung:2019ytz}, does not improve the $\chi^2$ nor does change the final extrapolation significantly.

The final extrapolation to the chiral and continuum limit of the strong coupling determined from the Wilson flow is presented in Fig.~\ref{fig:continuum_running_coupling}. There are no evidence of an infrared fixed point in the running of the strong coupling, nor a signal of a slowing of the growth of the strong coupling constant in the infrared regime. Our present calculation is in agreement with our previous study of the $\beta$-function in the Mini-MOM scheme, where no evidence of an infrared fixed point have been found in the region of momenta we have been able to explore \cite{Bergner:2017ytp}. Further studies close to the continuum limit and deeper in the infrared region will be required to confirm this result.

\section{Conclusion}

We have presented a numerical investigation of the $N_f=1$ AdjQCD theory using overlap fermions. Our results do not show any evidence of an infrared fixed point in the running of the gradient flow coupling in the region of energies that we have been able to explore. A non-zero extrapolated value of the scale $t_0$ indicates consistently that the theory doesn't show an infrared conformal behaviour.

Our results also support the breaking of chiral symmetry induced by a non-vanishing expectation value of the chiral condensate. Consequently, we would predict two light pions emerging as Goldstone bosons in the massless limit. In the previous investigations of Ref.~\cite{Georg:2015}, it has been observed that the glueball $0^+$ is rather the lightest particle in the spectrum. However, these investigations have been done in parameter regions significantly different from our study.

Ref.~\cite{Georg:2015} considers Wilson fermions, which allows a larger statistic and a determination of the particle spectrum. In addition rather fine lattices can be simulated with this fermion action. On the other hand, our current study considers the overlap operator which implements chiral symmetry on the lattice, but our lattices are rather coarse and we have not been able to determine particle masses. In the near future we plan to extend our current investigations to study of the bound state spectrum, in order to be able to identify directly whether pions become lighter than the glueball states.

{\bf Acknowledgements:} We thank G.~Münster for helpful comments. G.~B.\ and I.~S.\ acknowledge support from the Deutsche Forschungsgemeinschaft (DFG) Grant No.~BE 5942/3-1 and 5942/4-1.
The authors gratefully acknowledge the Gauss Centre for Supercomputing e.~V.\ (www.gauss-centre.eu) for funding this project by providing computing time on the GCS Supercomputer SuperMUC-NG at Leibniz Supercomputing Centre (www.lrz.de). Further computing time has been provided on the compute cluster PALMA
of the University of M\"unster and resources of Friedrich Schiller University Jena supported in part by DFG grants INST 275/334-1 FUGG and INST 275/363-1 FUGG. 

\bibliographystyle{utphys}
\bibliography{refs}

\providecommand{\href}[2]{#2}\begingroup\raggedright\begin{thebibliography}{10}

\bibitem{SEIBERG199419}
N.~Seiberg and E.~Witten, ``{Electric - magnetic duality, monopole
  condensation, and confinement in N=2 supersymmetric Yang-Mills theory},''
  \href{http://dx.doi.org/10.1016/0550-3213(94)90124-4}{{\em Nucl. Phys. B}
  {\bfseries 426} (1994) 19--52},
  \href{http://arxiv.org/abs/hep-th/9407087}{{\ttfamily arXiv:hep-th/9407087}}.
  [Erratum: Nucl.Phys.B 430, 485--486 (1994)].

\bibitem{Seiberg:1994pq}
N.~Seiberg, ``{Electric - magnetic duality in supersymmetric nonAbelian gauge
  theories},'' \href{http://dx.doi.org/10.1016/0550-3213(94)00023-8}{{\em Nucl.
  Phys. B} {\bfseries 435} (1995) 129--146},
  \href{http://arxiv.org/abs/hep-th/9411149}{{\ttfamily arXiv:hep-th/9411149}}.

\bibitem{Azeyanagi:2010ne}
T.~Azeyanagi, M.~Hanada, M.~Unsal, and R.~Yacoby, ``{Large-N reduction in
  QCD-like theories with massive adjoint fermions},''
  \href{http://dx.doi.org/10.1103/PhysRevD.82.125013}{{\em Phys. Rev. D}
  {\bfseries 82} (2010) 125013},
  \href{http://arxiv.org/abs/1006.0717}{{\ttfamily arXiv:1006.0717 [hep-th]}}.

\bibitem{Unsal:2010qh}
M.~Unsal and L.~G. Yaffe, ``{Large-N volume independence in conformal and
  confining gauge theories},''
  \href{http://dx.doi.org/10.1007/JHEP08(2010)030}{{\em JHEP} {\bfseries 08}
  (2010) 030}, \href{http://arxiv.org/abs/1006.2101}{{\ttfamily arXiv:1006.2101
  [hep-th]}}.

\bibitem{Bergner:2015adz}
G.~Bergner, P.~Giudice, G.~M\"unster, I.~Montvay, and S.~Piemonte, ``{The light
  bound states of supersymmetric SU(2) Yang-Mills theory},''
  \href{http://dx.doi.org/10.1007/JHEP03(2016)080}{{\em JHEP} {\bfseries 03}
  (2016) 080}, \href{http://arxiv.org/abs/1512.07014}{{\ttfamily
  arXiv:1512.07014 [hep-lat]}}.

\bibitem{Ali:2018dnd}
S.~Ali, G.~Bergner, H.~Gerber, P.~Giudice, I.~Montvay, G.~M\"unster,
  S.~Piemonte, and P.~Scior, ``{The light bound states of $\mathcal{N}=1$
  supersymmetric SU(3) Yang-Mills theory on the lattice},''
  \href{http://dx.doi.org/10.1007/JHEP03(2018)113}{{\em JHEP} {\bfseries 03}
  (2018) 113}, \href{http://arxiv.org/abs/1801.08062}{{\ttfamily
  arXiv:1801.08062 [hep-lat]}}.

\bibitem{bergner_spectrum_2017}
G.~Bergner, P.~Giudice, G.~M\"unster, I.~Montvay, and S.~Piemonte, ``{Spectrum
  and mass anomalous dimension of SU(2) adjoint QCD with two Dirac flavors},''
  \href{http://dx.doi.org/10.1103/PhysRevD.96.034504}{{\em Phys. Rev. D}
  {\bfseries 96} no.~3, (2017) 034504},
  \href{http://arxiv.org/abs/1610.01576}{{\ttfamily arXiv:1610.01576
  [hep-lat]}}.

\bibitem{bergner_low_2018}
G.~Bergner, P.~Giudice, G.~M\"unster, P.~Scior, I.~Montvay, and S.~Piemonte,
  ``{Low energy properties of SU(2) gauge theory with N$_{f}$ = 3/2 flavours of
  adjoint fermions},'' \href{http://dx.doi.org/10.1007/JHEP01(2018)119}{{\em
  JHEP} {\bfseries 01} (2018) 119},
  \href{http://arxiv.org/abs/1712.04692}{{\ttfamily arXiv:1712.04692
  [hep-lat]}}.

\bibitem{DelDebbio:2009fd}
L.~Del~Debbio, B.~Lucini, A.~Patella, C.~Pica, and A.~Rago, ``{Conformal versus
  confining scenario in SU(2) with adjoint fermions},''
  \href{http://dx.doi.org/10.1103/PhysRevD.80.074507}{{\em Phys. Rev. D}
  {\bfseries 80} (2009) 074507},
  \href{http://arxiv.org/abs/0907.3896}{{\ttfamily arXiv:0907.3896 [hep-lat]}}.

\bibitem{Hietanen:2009az}
A.~J. Hietanen, K.~Rummukainen, and K.~Tuominen, ``{Evolution of the coupling
  constant in SU(2) lattice gauge theory with two adjoint fermions},''
  \href{http://dx.doi.org/10.1103/PhysRevD.80.094504}{{\em Phys. Rev. D}
  {\bfseries 80} (2009) 094504},
  \href{http://arxiv.org/abs/0904.0864}{{\ttfamily arXiv:0904.0864 [hep-lat]}}.

\bibitem{Georg:2015}
A.~Athenodorou, E.~Bennett, G.~Bergner, and B.~Lucini, ``{Infrared regime of
  SU(2) with one adjoint Dirac flavor},''
  \href{http://dx.doi.org/10.1103/PhysRevD.91.114508}{{\em Phys. Rev. D}
  {\bfseries 91} no.~11, (2015) 114508},
  \href{http://arxiv.org/abs/1412.5994}{{\ttfamily arXiv:1412.5994 [hep-lat]}}.

\bibitem{Athenodorou:2021wom}
A.~Athenodorou, Bennett, G.~Bergner, and B.~Lucini, ``{Investigating the
  conformal behavior of SU(2) with one adjoint Dirac flavor},''
  \href{http://dx.doi.org/10.1103/PhysRevD.104.074519}{{\em Phys. Rev. D}
  {\bfseries 104} no.~7, (2021) 074519},
  \href{http://arxiv.org/abs/2103.10485}{{\ttfamily arXiv:2103.10485
  [hep-lat]}}.

\bibitem{Bi:2019gle}
Z.~Bi, A.~Grebe, G.~Kanwar, P.~Ledwith, D.~Murphy, and M.~L. Wagman, ``{Lattice
  Analysis of $SU(2)$ with 1 Adjoint Dirac Flavor},''
  \href{http://dx.doi.org/10.22323/1.363.0127}{{\em PoS} {\bfseries
  LATTICE2019} (2019) 127}, \href{http://arxiv.org/abs/1912.11723}{{\ttfamily
  arXiv:1912.11723 [hep-lat]}}.

\bibitem{Anber:2018iof}
M.~M. Anber and E.~Poppitz, ``{Two-flavor adjoint QCD},''
  \href{http://dx.doi.org/10.1103/PhysRevD.98.034026}{{\em Phys. Rev. D}
  {\bfseries 98} no.~3, (2018) 034026},
  \href{http://arxiv.org/abs/1805.12290}{{\ttfamily arXiv:1805.12290
  [hep-th]}}.

\bibitem{Poppitz:2021cxe}
E.~Poppitz, ``{Notes on Confinement on R3 \texttimes{} S1: From
  Yang\textendash{}Mills, Super-Yang\textendash{}Mills, and QCD (adj) to
  QCD(F)},'' \href{http://dx.doi.org/10.3390/sym14010180}{{\em Symmetry}
  {\bfseries 14} no.~1, (2022) 180},
  \href{http://arxiv.org/abs/2111.10423}{{\ttfamily arXiv:2111.10423
  [hep-th]}}.

\bibitem{Nielsen:1981hk}
H.~B. Nielsen and M.~Ninomiya, ``{No Go Theorem for Regularizing Chiral
  Fermions},'' \href{http://dx.doi.org/10.1016/0370-2693(81)91026-1}{{\em Phys.
  Lett. B} {\bfseries 105} (1981) 219--223}.

\bibitem{Ginsparg-Wilson}
P.~H. Ginsparg and K.~G. Wilson, ``{A Remnant of Chiral Symmetry on the
  Lattice},'' \href{http://dx.doi.org/10.1103/PhysRevD.25.2649}{{\em Phys. Rev.
  D} {\bfseries 25} (1982) 2649}.

\bibitem{Cundy:2005pi}
N.~Cundy, S.~Krieg, G.~Arnold, A.~Frommer, T.~Lippert, and K.~Schilling,
  ``{Numerical methods for the QCD overlap operator IV: Hybrid Monte Carlo},''
  \href{http://dx.doi.org/10.1016/j.cpc.2008.08.006}{{\em Comput. Phys.
  Commun.} {\bfseries 180} (2009) 26--54},
  \href{http://arxiv.org/abs/hep-lat/0502007}{{\ttfamily
  arXiv:hep-lat/0502007}}.

\bibitem{Wenger:2006ps}
U.~Wenger, ``{Chiral fermions on the lattice: A Flatlander's ascent into five
  dimensions},'' \href{http://dx.doi.org/10.1016/j.nima.2005.11.153}{{\em Nucl.
  Instrum. Meth. A} {\bfseries 559} (2006) 237--241}.

\bibitem{piemonte_monte-carlo_2020}
S.~Piemonte, G.~Bergner, and C.~L\'opez, ``{Monte Carlo simulations of overlap
  Majorana fermions},''
  \href{http://dx.doi.org/10.1103/PhysRevD.102.014503}{{\em Phys. Rev. D}
  {\bfseries 102} no.~1, (2020) 014503},
  \href{http://arxiv.org/abs/2005.02236}{{\ttfamily arXiv:2005.02236
  [hep-lat]}}.

\bibitem{Unsal:2008eg}
M.~Unsal, ``{Quantum phase transitions and new scales in QCD-like theories},''
  \href{http://dx.doi.org/10.1103/PhysRevLett.102.182002}{{\em Phys. Rev.
  Lett.} {\bfseries 102} (2009) 182002},
  \href{http://arxiv.org/abs/0807.0466}{{\ttfamily arXiv:0807.0466 [hep-th]}}.

\bibitem{Creutz:2006ts}
M.~Creutz, ``{One flavor QCD},''
  \href{http://dx.doi.org/10.1016/j.aop.2007.01.002}{{\em Annals Phys.}
  {\bfseries 322} (2007) 1518--1540},
  \href{http://arxiv.org/abs/hep-th/0609187}{{\ttfamily arXiv:hep-th/0609187}}.

\bibitem{Luscher:2009eq}
M.~Luscher, ``{Trivializing maps, the Wilson flow and the HMC algorithm},''
  \href{http://dx.doi.org/10.1007/s00220-009-0953-7}{{\em Commun. Math. Phys.}
  {\bfseries 293} (2010) 899--919},
  \href{http://arxiv.org/abs/0907.5491}{{\ttfamily arXiv:0907.5491 [hep-lat]}}.

\bibitem{Luscher:2010iy}
M.~L\"uscher, ``{Properties and uses of the Wilson flow in lattice QCD},''
  \href{http://dx.doi.org/10.1007/JHEP08(2010)071}{{\em JHEP} {\bfseries 08}
  (2010) 071}, \href{http://arxiv.org/abs/1006.4518}{{\ttfamily arXiv:1006.4518
  [hep-lat]}}. [Erratum: JHEP 03, 092 (2014)].

\bibitem{Luscher:2011bx}
M.~Luscher and P.~Weisz, ``{Perturbative analysis of the gradient flow in
  non-abelian gauge theories},''
  \href{http://dx.doi.org/10.1007/JHEP02(2011)051}{{\em JHEP} {\bfseries 02}
  (2011) 051}, \href{http://arxiv.org/abs/1101.0963}{{\ttfamily arXiv:1101.0963
  [hep-th]}}.

\bibitem{Hasenfratz:2019hpg}
A.~Hasenfratz and O.~Witzel, ``{Continuous renormalization group $\beta$
  function from lattice simulations},''
  \href{http://dx.doi.org/10.1103/PhysRevD.101.034514}{{\em Phys. Rev. D}
  {\bfseries 101} no.~3, (2020) 034514},
  \href{http://arxiv.org/abs/1910.06408}{{\ttfamily arXiv:1910.06408
  [hep-lat]}}.

\bibitem{Husung:2019ytz}
N.~Husung, P.~Marquard, and R.~Sommer, ``{Asymptotic behavior of cutoff effects
  in Yang\textendash{}Mills theory and in Wilson\textquoteright{}s lattice
  QCD},'' \href{http://dx.doi.org/10.1140/epjc/s10052-020-7685-4}{{\em Eur.
  Phys. J. C} {\bfseries 80} no.~3, (2020) 200},
  \href{http://arxiv.org/abs/1912.08498}{{\ttfamily arXiv:1912.08498
  [hep-lat]}}.

\bibitem{Bergner:2017ytp}
G.~Bergner and S.~Piemonte, ``{Running coupling from gluon and ghost
  propagators in the Landau gauge: Yang-Mills theories with adjoint
  fermions},'' \href{http://dx.doi.org/10.1103/PhysRevD.97.074510}{{\em Phys.
  Rev. D} {\bfseries 97} no.~7, (2018) 074510},
  \href{http://arxiv.org/abs/1709.07367}{{\ttfamily arXiv:1709.07367
  [hep-lat]}}.

\end{thebibliography}\endgroup

\end{document}